\documentclass[journal]{IEEEtran}
\usepackage{amssymb,bbm,amsmath,color,psfrag,subfigure}
\usepackage[draft]{graphicx}
\usepackage{xspace,stackrel,hyperref}
\usepackage{wrapfig}
\usepackage{mathtools}
\usepackage[ruled,norelsize]{algorithm2e}

\SetKwComment{Comment}{$\triangleright$\ }{}

\SetCommentSty{mycommfont}

\usepackage{subfigure}
\usepackage{tikz,pgfplots}
\usepackage{tikzexternal}

\tikzsetexternalprefix{./tikzcache/}

\def\beq{\begin{equation}}
\def\eeq{\end{equation}}

\newcommand{\ds}{\displaystyle}

\newcommand{\ba}{\begin{array}}
\newcommand{\ea}{\end{array}}

\newcommand{\be}{\begin{equation}}
\newcommand{\ee}{\end{equation}}

\newcommand{\1}{\mathbbm{1}}

\newcommand{\R}{\mathbb{R}}

\newcommand{\se}{\text{ if }}

\DeclareMathOperator*{\argmax}{argmax}

\newcommand{\mc}{\mathcal}

\newcommand*{\optMinMax}[2]{\underset{\vphantom{A}\displaystyle\mathclap{#1}}%
                           {\text{#2}}\quad}

\newcommand{\optmax}[1]{\optMinMax{#1}{maximize}}
\newcommand*{\subject}{\text{subject to}\quad}

\newcommand{\RR}{\mathbb{R}}

\newtheorem{remark}{Remark}

\newtheorem{example}{Example}

\def\QEDclosed{\mbox{\rule[0pt]{1.3ex}{1.3ex}}} % for a filled box
\def\QED{\QEDclosed} % default to closed

\newcommand\oprocendsymbol{\hbox{$\square$}}
\newcommand\oprocend{\relax\ifmmode\else\unskip\hfill\fi\oprocendsymbol}

\renewcommand{\theenumi}{(\roman{enumi})}

\makeatletter
\newcommand{\removelatexerror}{\let\@latex@error\@gobble}
\makeatother

\definecolor{mycolor1}{RGB}{230,97,1}
\definecolor{mycolor2}{RGB}{253,184,99}
\definecolor{mycolor3}{RGB}{178,171,210}
\definecolor{mycolor4}{RGB}{94,60,153}

\definecolor{mygreen}{rgb}{0,0.8,0}
\definecolor{myred}{rgb}{1,0,0}
\definecolor{myyellow}{rgb}{0.9, 0.9,0}

\begin{document}

\title{A Micro-Simulation Study of the Generalized Proportional Allocation Traffic Signal Control} 

\author{Gustav Nilsson\thanks{G.~Nilsson is with the Department of Automatic Control, Lund University, Sweden. \texttt{gustav.nilsson@control.lth.se}} and  Giacomo Como\thanks{G.~Como is with the Department of Mathematical Sciences, Politecnico di Torino, Italy, and the Department of Automatic Control, Lund University, Sweden. \texttt{giacomo.como@polito.it}}\thanks{The authors are members of the excellence centers LCCC and ELLIT. This reasearch was carried on within the framework of the MIUR-funded {\it Progetto di Eccellenza} of the {\it Dipartimento di Scienze Matematiche G.L.~Lagrange}, CUP: E11G18000350001, and was partly supported by the {\it Compagnia di San Paolo} and the Swedish Research Council (VR).}

}

\maketitle

\begin{abstract}
In this paper, we study the problem of determining phase activations for signalized junctions by utilizing feedback, more specifically, by measure the queue-lengths on the incoming lanes to each junction. The controller we are investigating is the Generalized Proportional Allocation (GPA) controller, which has previously been shown to have desired stability and throughput properties in a continuous averaged dynamical model for queueing networks. In this paper, we provide and implement two discretized versions of the GPA controller in the SUMO micro simulator. We also compare the GPA controllers with the MaxPressure controller, a controller that requires more information than the GPA, in an artificial Manhattan-like grid. To show that the GPA controller is easy to implement in a real scenario, we also implement it in a previously published realistic traffic scenario for the city of Luxembourg and compare its performance with the static controller provided with the scenario. The simulations show that the GPA performs better than a static controller for the Luxembourg scenario, and better than the MaxPressure pressure controller in the Manhattan-grid when the demands are low.
\end{abstract}

\textbf{Index terms:} Decentralized Traffic Signal Control, Microscopic Traffic Simulation

\section{Introduction}

While the first traffic signals were controlled completely in open loop, various approaches have been taken to adjust the green light allocation based on the current traffic situation. To mention a few, SCOOT~\cite{robertson1991optimizing}, UTOPIA~\cite{mauro1990utopia} and SCATS~\cite{sims1980sydney}. Also, learning based approaches have been taken, e.g.,~\cite{JIN20175301}.

However, these approaches lack of formal stability, optimality, and robustness guarantees. In~\cite{nilsson2015entropy, nilsson2017generalized}, a decentralized feedback controller for traffic control was proposed, refereed to as Generalized Proportional Allocation (GPA) controller, which has both stability and maximal throughput guarantees. In those papers, an average control action for traffic signals in continuous time is given.  Since the controller has several desired properties, it is motivated to investigate if this controller performs well in a micro-simulator with more realistic traffic dynamics. First of all, under the assumptions that the controller can measure the whole queue lengths at each junction, the averaged controller is throughput optimal from a theoretical perspective. With this, we mean that when the traffic dynamics is modeled as a simple system of point queues there exists no controller that can handle larger constant exogenous inflows to a network than this controller. This property of throughput-optimality also means that there are formal guarantees that the controller will not create gridlock situations in the network. As exemplified in~\cite{varaiya2013max}, feedback controllers that perform well for a single isolated junction may cause gridlock situations in a network setting.

At the same time, this controller requires very little information about the network topology and the traffic flow propagation. All information the controller needs to determine the phase activation in a junction is the queue lengths on the incoming lanes to a junction and the static set of phases.  Those requirements on information make the controller fully distributed, i.e., to compute the control action in one junction, no information is required about the state in the other junctions.

The proposed traffic signal controller also has the property that it adjusts the cycle lengths depending on the demand. The fact that during higher demands, the cycle lengths should be longer to waste less service time due to phase shifts, has been suggested previously for open loop traffic signal control, see e..g~\cite{roess2011traffic}.

Another feedback control strategy for traffic signal control is the MaxPressure controller~\cite{Varaiya:13, varaiya2013max}. The MaxPressure controller utilizes the same idea as the BackPressure controller, proposed for communication networks in~\cite{tassiulas1992stability}. While the BackPressure controller controls both the routing (to which packets the should proceed after received service)  and the scheduling (which subset of queues that should be severed), the MaxPressure controller only controls the latter, i.e., the phase activation but not the routing. More recently, due to the rapid development of autonomous vehicles, it has been proposed in~\cite{zaidi2018backpressure} to utilize the routing control from the BackPressure controller in traffic networks as well. The MaxPressure controller is also throughput optimal, but it requires information about the tuning ratios at each junction, i.e., how the vehicles (in average) propagate from one junction to the neighboring junctions. Although various techniques for estimating those turning ratios have been made, for example~\cite{coogan2017traffic}, with more and more drivers or autonomous vehicles doing their path planning through some routing service, it is likely to believe that the turning ratios can change in an unpredictable way when a disturbance occurs in the traffic network. 

If the traffic signal controller has information about the turning ratios, other control strategies are possible as well, for instance, MPC-like as proposed in~\cite{hao2018modelI, hao2018modelII, grandinetti2018distributed} and robust control as proposed in~\cite{bianchin2018network}.  

In~\cite{nilsson2018} we presented the first discretization and validation results of the GPA in a microscopic traffic simulator. Although, the results were promising, the validations were only performed on an artificial network and only compared with a fixed timed traffic signal controller. Moreover, the GPA was only discretized in a way such that the full cycle is activated. In this paper, we extend the results in~\cite{nilsson2018} by showing another discretization that does not have to utilize the full cycle and we also perform new validations. The new validations both compare the GPA to the MaxPressure controller on an artificial network (the reason for chosen a artificial network will be explained later), but also validate the GPA controller in a realistic scenario, namely for the Luxembourg city during a whole day.

The outline of the paper is as follows: In Section~\ref{sec:problem} we present the model we are using for traffic signals, together with a problem formulation of the traffic signal control problem. In Section~\ref{sec:controllers} we present two different discretization of the GPA that we are using in this study, but also give a brief description of the MaxPressure controller. In Section~\ref{sec:comparision} we compare the GPA controller with the MaxPressure controller on an artificial Manhattan-like grid, and in Section~\ref{sec:lust} we investigate how the GPA controller performs in a realistic traffic scenario. The paper is concluded with some ideas about further research.

\subsection{Notation}
We let $\RR_+$ denote the non-negative reals. For a finite sets $\mc A, \mc B$, we let $\RR_+^{\mc A}$ denote non-negative vectors indexed by the elements in $\mc A$, and $\RR_+^{\mc A \times \mc B}$ the matrices indexed by elements $\mc A$ and $\mc B$.

\section{Model and Problem Formulation}\label{sec:problem}
In this section, we describe the model for traffic signals to be used throughout the paper together with the associated control problem.

We consider an arterial traffic network with signalized junctions. Let $\mc J$ denote the set of signalized junctions. For a junction $j \in \mc J$, we let $\mc L^{(j)}$ be the set of incoming lanes, on which the vehicles can queue up. The set of all signalized lanes in the whole network will be denoted by $\mc L = \cup_{j \in \mc J} \mc L^{(j)}$. For a lane $l \in \mc L^{(j)}$, the queue-length at time $t$ --measured in the number of vehicles--  is denoted by $x_l(t)$.

Each junction has a predefined set of \emph{phases} $\mc P^{(j)}$ of size $n_{p_j}$. For simplicity, we assume that phases $p_i \in \mc P^{(j)}$ are indexed by $i = 1, \ldots, n_{p_j}$. A phase  $p \in \mc P^{(j)}$ is a subset of incoming lanes to the junction $j$ that can receive green light simultaneously. Throughout the paper, we will assume that for each lane $l \in \mc L$, there exists only one junction $j \in \mc J$ and at least one phase $p \in \mc P^{(j)}$ such that $l \in p$.

The phases are usually constructed such that the vehicles paths in a junction do not cross each other. This to avoid collisions. %, even though sometimes the collision avoidance is sometimes the driver's responsibility. 
Examples of this will be shown later in this paper. After a phase has been activated, it is common to signalize to the drivers that the traffic signal is turning red and give time for vehicles that are in the middle of the junction to leave it before the next phase are activated. Such time is usually referred to as clearance time. Throughout the paper we shall refer to those phases only containing red and yellow traffic light as \emph{clearance phases} (in contrast to phases, that models when lanes receives green traffic light). We will assume that each phase activation is followed by a clearance phase activation. While we will let the phase activation time vary, we will make the quite natural assumption that the clearance phases has to be activated for a fixed time.

For a given junction $j \in \mc J$, the set of phases can be described through a phase matrix $P^{(j)}$, where 
$$P_{il}=\left\{\ba{lcl}1&\se&\text{lane }l\text{ belongs to the }i\text{-th phase}\\ 0&\se&\text{otherwise\,.}\ea\right.$$

While the phase matrix does not contain the clearance phases, to each phase $p \in \mc P^{(j)}$ we will associate a clearance phase, denoted $p'$. We denote the set of real phases and their corresponding clearance phases $\bar{\mc P}^{(j)}$.

The controller's task in a signalized junction is to define a \emph{signal program}, $\mc T^{(j)} = \{ (p, t_\text{end} ) \in  \bar{\mc P}^{(j)} \times  \R_+  \}$, where the phase $p$ is activated until $t_\text{end}$. When $t = t_\text{end}$, the phase $p'$, where $(p', t_\text{end}) \in \mc T^{(j)}$, with smallest $t_\text{end} > t$ is activated. Formally, we can define the function $c^{(j)}(t)$ that gives the phase that is activated at time $t$ as follows
\begin{align*}
c^{(j)} (t) =  \{ & p : (p, t_\text{end}) \in { \mc T}^{(j)}  \mid \\ &  t_\text{end} > t \text{ and }  t_\text{end} \leq t'_\text{end} \textrm{ for all } (p', t'_\text{end}) \in   { \mc T}^{(j)}  \} \, .
\end{align*}
What $c^{(j)} (t)$ is doing is to find the phase with the smallest end-time greater than the current time.

\medskip
\begin{example} \label{ex:phasesandprogram}
\begin{figure}
\centering
\begin{tikzpicture}
\begin{scope}[scale=0.5]
\draw[thick] (-3, 1) -- (-1, 1) -- (-1,3);
\draw[thick] (-3, -1) -- (-1, -1) -- (-1, -3);
\draw[thick] (1,3) -- (1,1) -- (3,1);
\draw[thick] (1, -3) -- (1,-1) -- (3, -1);

\draw [->, thick] (-1, -0.5) to [bend right] (0.3, 1);
\draw [->, thick] (-1, -0.5) to (1, -0.5);
\draw [->, thick] (-1, -0.5) to [bend left] (-0.7, -1);

\draw [->, thick] (1, 0.5) to [bend left] (0.7, 1);
\draw [->, thick] (1, 0.5) to (-1, 0.5);
\draw [->, thick] (1, 0.5) to [bend right] (-0.3, -1);

\node (l1) at (-1.5, -0.5) {$l_1$};
\node (l2) at (0.5, -1.5) {$l_2$};
\node (l3) at (1.5, 0.5) {$l_3$};
\node (l4) at (-0.5, 1.5) {$l_4$};
\end{scope}

\begin{scope}[scale=0.5, shift={(7, 0)}]
\begin{scope}[rotate=90]
\draw[thick] (-3, 1) -- (-1, 1) -- (-1,3);
\draw[thick] (-3, -1) -- (-1, -1) -- (-1, -3);
\draw[thick] (1,3) -- (1,1) -- (3,1);
\draw[thick] (1, -3) -- (1,-1) -- (3, -1);

\draw [->, thick] (-1, -0.5) to [bend right] (0.3, 1);
\draw [->, thick] (-1, -0.5) to (1, -0.5);
\draw [->, thick] (-1, -0.5) to [bend left] (-0.7, -1);

\draw [->, thick] (1, 0.5) to [bend left] (0.7, 1);
\draw [->, thick] (1, 0.5) to (-1, 0.5);
\draw [->, thick] (1, 0.5) to [bend right] (-0.3, -1);
\draw[dashed] (0, 3) -- (0,1);
\draw[dashed] (-3, 0) -- (-1, 0);
\draw[dashed] (0, -1) -- (0, -3);
\draw[dashed] (3, 0) -- (1, 0);
\end{scope}
\node (l1) at (-1.5, -0.5) {$l_1$};
\node (l2) at (0.5, -1.5) {$l_2$};
\node (l3) at (1.5, 0.5) {$l_3$};
\node (l4) at (-0.5, 1.5) {$l_4$};
\end{scope}

\begin{scope}[scale=0.5]
\draw[dashed] (0, 3) -- (0,1);
\draw[dashed] (-3, 0) -- (-1, 0);
\draw[dashed] (0, -1) -- (0, -3);
\draw[dashed] (3, 0) -- (1, 0);
\end{scope}
\end{tikzpicture}
\caption{The phases for the junction in Example~\ref{ex:phasesandprogram}. This junction has four incoming lanes and two phases, $p_1 = \{l_1, l_3\}$ and $p_2 = \{l_2, l_4\}$. Hence there is no specific lane left-turning left.}
\label{fig:phasesexamplejunc}
\end{figure}
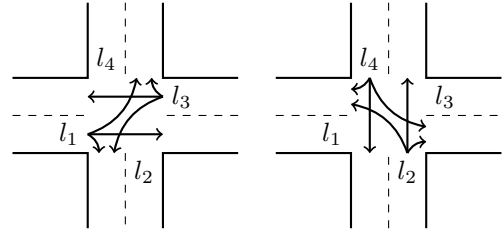
\begin{figure}
\centering
\begin{tikzpicture}[scale=1.2]
\draw[->] (0, 0) -- (6.5, 0) node[right] {$t$};

\draw[-] (0, 0.1) -- (0, -0.1) node[below] {$0$} ;
\draw[-] (2.5, 0.1) -- (2.5, -0.1) node[below] {$25$} ; %p1
\draw[-] (3, 0.1) -- (3, -0.1) node[below] {$30$} ; %p1'
\draw[-] (5.5, 0.1) -- (5.5, -0.1) node[below] {$55$} ; %p2
\draw[-] (6, 0.1) -- (6, -0.1) node[below] {$60$}; %p2'

\node (c) at (0, 0.4) {$c(t)$};
\node (p1) at (1.25, 0.4) {$p_1$};
\node (p1p) at (2.75, 0.4) {$ p_1'$};
\node (p2) at (4.25, 0.4) {$p_2$};
\node (p2p) at (5.75, 0.4) {$ p_2'$};

\begin{scope}[scale=0.20, shift={(6, 8)}]
\draw[thick] (-3, 1) -- (-1, 1) -- (-1,3);
\draw[thick] (-3, -1) -- (-1, -1) -- (-1, -3);
\draw[thick] (1,3) -- (1,1) -- (3,1);
\draw[thick] (1, -3) -- (1,-1) -- (3, -1);

\fill[mygreen] (-1, -1) -- (-1.4, -1) -- (-1.4, 0) -- (-1, 0) -- cycle;
\fill[mygreen] (1, 1) -- (1.4, 1) -- (1.4, 0) -- (1, 0) -- cycle;
\fill[myred] (1, -1) -- (1, -1.4) -- (0, -1.4) -- (0, -1) -- cycle;
\fill[myred] (-1, 1) -- (-1,1.4) -- (0, 1.4) -- (0, 1) -- cycle;
\end{scope}

\begin{scope}[scale=0.20, shift={(13.75, 8)}]
\draw[thick] (-3, 1) -- (-1, 1) -- (-1,3);
\draw[thick] (-3, -1) -- (-1, -1) -- (-1, -3);
\draw[thick] (1,3) -- (1,1) -- (3,1);
\draw[thick] (1, -3) -- (1,-1) -- (3, -1);

\fill[myyellow] (-1, -1) -- (-1.4, -1) -- (-1.4, 0) -- (-1, 0) -- cycle;
\fill[myyellow] (1, 1) -- (1.4, 1) -- (1.4, 0) -- (1, 0) -- cycle;
\fill[myred] (1, -1) -- (1, -1.4) -- (0, -1.4) -- (0, -1) -- cycle;
\fill[myred] (-1, 1) -- (-1,1.4) -- (0, 1.4) -- (0, 1) -- cycle;
\end{scope}

\begin{scope}[scale=0.20, shift={(21.25, 8)}]
\draw[thick] (-3, 1) -- (-1, 1) -- (-1,3);
\draw[thick] (-3, -1) -- (-1, -1) -- (-1, -3);
\draw[thick] (1,3) -- (1,1) -- (3,1);
\draw[thick] (1, -3) -- (1,-1) -- (3, -1);

\fill[myred] (-1, -1) -- (-1.4, -1) -- (-1.4, 0) -- (-1, 0) -- cycle;
\fill[myred] (1, 1) -- (1.4, 1) -- (1.4, 0) -- (1, 0) -- cycle;
\fill[mygreen] (1, -1) -- (1, -1.4) -- (0, -1.4) -- (0, -1) -- cycle;
\fill[mygreen] (-1, 1) -- (-1,1.4) -- (0, 1.4) -- (0, 1) -- cycle;
\end{scope}

\begin{scope}[scale=0.20, shift={(28.75, 8)}]
\draw[thick] (-3, 1) -- (-1, 1) -- (-1,3);
\draw[thick] (-3, -1) -- (-1, -1) -- (-1, -3);
\draw[thick] (1,3) -- (1,1) -- (3,1);
\draw[thick] (1, -3) -- (1,-1) -- (3, -1);

\fill[myred] (-1, -1) -- (-1.4, -1) -- (-1.4, 0) -- (-1, 0) -- cycle;
\fill[myred] (1, 1) -- (1.4, 1) -- (1.4, 0) -- (1, 0) -- cycle;
\fill[myyellow] (1, -1) -- (1, -1.4) -- (0, -1.4) -- (0, -1) -- cycle;
\fill[myyellow] (-1, 1) -- (-1,1.4) -- (0, 1.4) -- (0, 1) -- cycle;
\end{scope}

% All the dashed lines, MC fulhaxx
\begin{scope}[scale=0.20, shift={(6, 8)}]
\draw[dashed] (0, 3) -- (0,1);
\draw[dashed] (-3, 0) -- (-1, 0);
\draw[dashed] (0, -1) -- (0, -3);
\draw[dashed] (3, 0) -- (1, 0);
\end{scope}

\begin{scope}[scale=0.20, shift={(13.75, 8)}]
\draw[dashed] (0, 3) -- (0,1);
\draw[dashed] (-3, 0) -- (-1, 0);
\draw[dashed] (0, -1) -- (0, -3);
\draw[dashed] (3, 0) -- (1, 0);
\end{scope}

\begin{scope}[scale=0.20, shift={(21.25, 8)}]
\draw[dashed] (0, 3) -- (0,1);
\draw[dashed] (-3, 0) -- (-1, 0);
\draw[dashed] (0, -1) -- (0, -3);
\draw[dashed] (3, 0) -- (1, 0);
\end{scope}

\begin{scope}[scale=0.20, shift={(28.75, 8)}]
\draw[dashed] (0, 3) -- (0,1);
\draw[dashed] (-3, 0) -- (-1, 0);
\draw[dashed] (0, -1) -- (0, -3);
\draw[dashed] (3, 0) -- (1, 0);
\end{scope}

\end{tikzpicture}
\caption{Example of a signal program for the junction in Example~\ref{ex:phasesandprogram}. In this example the signal program is $\mc T = \{ (p_1, 25), (p_1', 30), (p_2, 55), (p_2', 60)\}$.}
\label{fig:signaltiming}
\end{figure}
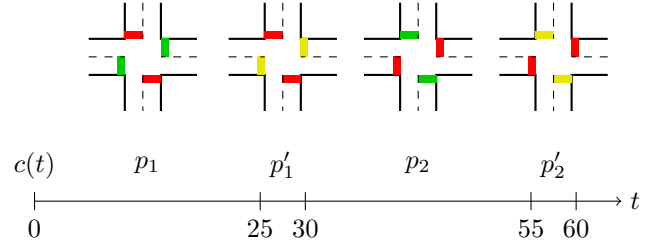
Consider the junction in Fig.~\ref{fig:phasesexamplejunc} with the incoming lanes numbered as in the figure. In this case the drivers turning left have to solve the collision avoidance by themselves. The phase matrix is
$$P = \begin{bmatrix} 1 & 0 & 1 & 0 \\ 0 & 1 & 0 & 1  \end{bmatrix} \, .$$

An example of signal program is shown in Fig.~\ref{fig:signaltiming}. Here the program is $\mc T = \{ (p_1, 25), (p_1', 30), (p_2, 55), (p_2', 60)\}$. which means that both the phases are activated for $25$ seconds each, and the clearance phases are activated for $5$ seconds each.
\end{example}
\medskip

Moreover, we let 
$$T^{(j)} = \max\{t_\text{end}  \mid (p, t_\text{end}) \in {\mc P}^{(j)} \}$$
denote the time when the signal program for junction $j$ ends, and hence a new signal timing program has to be determined.

\section{Feedback Controllers}\label{sec:controllers}
In this section, we present three different traffic signal controllers that all determine the signal program. The first two are discretization of the GPA controller, where the first one makes sure that all the clearance phase are activated during one cycle, and the second one only activates the clearance phases if their corresponding phase has been activated. The third controller is the MaxPressure controller. 

All the three controllers are feedback-based, i.e., when one signal program has reached its end, the current queue lengths are used to determine the upcoming signal program. Moreover, the GPA controllers are fully distributed, in the sense that to determine the signal program in one junction, the controller only needs information about the queue-lengths on the incoming lanes for that junction. The MaxPressure controller is also distributed in the sense that it does not requires network wide information, but it requires queue length information from the neighboring junctions as well.

For all of the controller presented in this section, we assume for simplicity of the presentation that after a phase has been activated, a clearance phase has to be activated for a fixed amount of time $T_w > 0$, that is independent of which phase that has just been activated.

\subsection{GPA with Full Clearance Cycles} \label{sec:GPAfull}
For this controller, we assume that all the clearance phases have to be activated for each cycle. When $t = T^{(j)}$, a new signal program is computed by solving the following convex optimization problem:
\begin{equation}\label{eq:gpa}
\begin{aligned}
\optmax{\begin{matrix} \hspace{0.2em} \nu\in\R_+^{n_{p_j}} \\  w\in\R_+ \end{matrix}} &  \sum_{l \in \mc L^{(j)}} x_l(t) \log\left( (P^T\nu)_l \right) + \kappa \log(w)  \, , \\
\subject &  \sum_{1 \leq i \leq n_{p_j}} \nu_i + w = 1 \,, \\
&  w \geq \bar{w} \, .
\end{aligned}
\end{equation}
In the optimization problem above, $\kappa > 0$ and $\bar{w} \geq 0$ are tuning parameters for the controller, and their interpretation will be discussed later.

The vector $\nu$ in the solution of the optimization problem above, determines the fraction of the cycle time that each phase should be activated, where each element in $\nu$ contains this fraction. The variable $w$ tells how large fraction of the cycle time that should be allocated to the clearance phases. Observe that as long as the queue lengths are finite $w$ will be strictly greater than zero. Since we assume that each clearance phase has to be activated for a fixed amount of time, $T_w > 0$, the total cycle length $T_\text{cyc}$ for the upcoming cycle can be computed by
$$T_\text{cyc} = \frac{n_{p_j} T_w}{w} \, .$$
With the knowledge of the full-cycle length, the signal program for the upcoming cycle can be computed according to Algorithm~\ref{algo:gpafull}.

Although the optimization problem can be solved in real-time using convex solvers, the optimization problem can also be solved analytically in the spacial cases. One such case is when the phases are orthogonal, i.e., every incoming lane only belongs to one phase. If the phases are orthogonal, then $P^T \1 = \1$. In the case of orthogonal phases and $\bar{w} = 0$, the solution to the optimization problem in~\eqref{eq:gpa} is given by

\begin{equation} \label{eq:gpaorthogonal}
\begin{aligned}
 \nu_i (x(t)) &= \frac{\sum_{l \in \mc L^{(j)}}P_{il}x_l(t)}{\kappa+\sum_{l \in \mc L^{(j)}} x_l(t)}\,,\qquad i=1,\ldots,n_{p_j}\,,  \\
 w(x(t)) &= \frac{\kappa}{\kappa+\sum_{l \in \mc L^{(j)}} x_l(t)} \,.
 \end{aligned}
\end{equation}
From the expression of $w$ above, a direct expression for the total cycle length can be obtained
\begin{equation*}
\ds T_\text{cyc} = T_w n_{p_j} +\frac{T_w n_{p_j}}{\kappa}{\sum_{l \in \mc L^{(j)}} x_l(t)} \, .\label{eq:cycletime}
\end{equation*}

From the expressions above we can observe a few things. First, we see that the fraction of the cycle that each phase is activated is proportional to the queue lengths in that phase, and this explains why we done this control strategy generalized proportional allocation. Moreover, we get an interpretation of the tuning parameter $\kappa$, it tells how the cycle length $T_\text{cyc}$ should scale with the current queue lengths. If $\kappa$ is small, even small queue lengths will cause longer cycles, while if $\kappa$ is large the cycles will be short even for large queues. Hence, a too small $\kappa$ may give too long cycles, which can result in that lanes get more green-light than needed and the controller ends up giving green light to empty lanes, while vehicles in other lanes are waiting for service. On the other hand, a too large $\kappa$ may make the cycle lengths so short, so that the fraction of the cycle that each phase gets activated is too short for the drivers to react on.

\begin{figure}[!t]
\removelatexerror
\begin{algorithm}[H]
\caption{GPA with Full Clearance Cycles}\label{algo:gpafull}
\DontPrintSemicolon
\KwData{Current time $t$, local queue lengths $x^{(j)}(t)$, phase matrix $P^{(j)}$, clearance time $T_w$, tuning parameters $\kappa, \bar w$}
\KwResult{Signal program $\mc T^{(j)}$}
%\Begin{
   $\mc T^{(j)} \leftarrow \emptyset$ \;
   $n_{p_j} \leftarrow $ Number of rows in $P^{(j)}$ \;
   $(\nu, w)$ $\leftarrow$ Solution to~\eqref{eq:gpa} given $x^{(j)}(t), P^{(j)}, \kappa, \bar w$\;
   $T_\text{cyc} \leftarrow n_p \cdot T_w / w$ \;
   $t_\text{end} \leftarrow t$ \;
  \For{$i\leftarrow 1$ \KwTo $n_{p_j}$}{
     	$t_\text{end} \leftarrow t_\text{end} + \nu_i \cdot T_\text{cyc}$ \;
	$\mc T^{(j)} \leftarrow \mc T^{(j)} + (p_i, t_\text{end})$  \Comment*[r]{Add phase $p_i$} 
	$t_\text{end} \leftarrow t_\text{end} + T_w$ \;
         $\mc T^{(j)} \leftarrow \mc T^{(j)} + (p'_i, t_\text{end})$ \Comment*[r]{Add clearance phase $p_i'$} 
   }
%}
\end{algorithm}
\end{figure}

\begin{remark}
In~\cite{nilsson2017generalized} we showed that the averaged continuous time GPA controller can stabilize, and hence keep the queue-lengths bounded, the network. Moreover, this averaged version is throughput-optimal, which means that no controller can handle more exogenous inflow to network than this controller. 
\end{remark}

However, when the controller is discretized, the following example shows that an upper bound on the cycle length, i.e., $\bar{w} > 0$ is required to guarantee stability even for an isolated junction.

\begin{example}\label{ex:unstableunboundedcycle}
Consider a junction with two incoming lanes with unit flow capacity, both having their own phase, and let the exogenous inflows $\lambda_1 = \lambda_2 = \lambda$, $T_w = 1$, $\bar w = 0$, $x_1(0) = A > 0$, and $x_2(0)  = 0$. The control signals and the cycle time for the first iteration is then given by
\begin{align*}
u_1(x(0)) &=  \frac{A}{A+\kappa} \, , \\
u_2(x(0)) &= 0 \, , \\ 
T(x(0)) &= \frac{A+\kappa}{\kappa}.
\end{align*}
Observe that the cycle time $T(x(0))$ is strictly increasing with $A$. After one full service cycle, i.e., at $t_1 = T(x(0))$ the queue lengths are
\begin{align*}
x_1(t_1) &= A + T(x(0)) \left(\lambda - \frac{A}{A+\kappa} \right)= \overbrace{A + \lambda \frac{A+\kappa}{\kappa} - \frac{A}{\kappa}}^{f(A)} \, , \\
x_2(t_1) &= T(x(0)) \lambda = \lambda \left( \frac{A+ \kappa}{\kappa} \right).
\end{align*}
If $x_1(t_1) = 0$, then due to symmetry, the analysis of the system can be repeated in the same way with a new initial condition. To make sure that one queue always get empty during the service cycles, it must hold that $f(A) \leq 0$. Moreover, to make sure that the other queue grows, it must also hold that $x_2(t_1) > A$ which can be equivalently expressed as
\begin{align*}
A \kappa  + \lambda(A + \kappa) - A &\leq 0 \, , \\
A \kappa - \lambda(A+\kappa) &< 0 \, .
\end{align*}
The choice of $\lambda = \kappa =  0.1$ and $A= 1$ is one set of parameters satisfying the constraints above, and will hence make the queue lengths and cycle times grow unboundedly. How queue lengths and cycle times evolve in this case is shown in Fig.~\ref{fig:unstableunboundedcycle}.
\end{example}
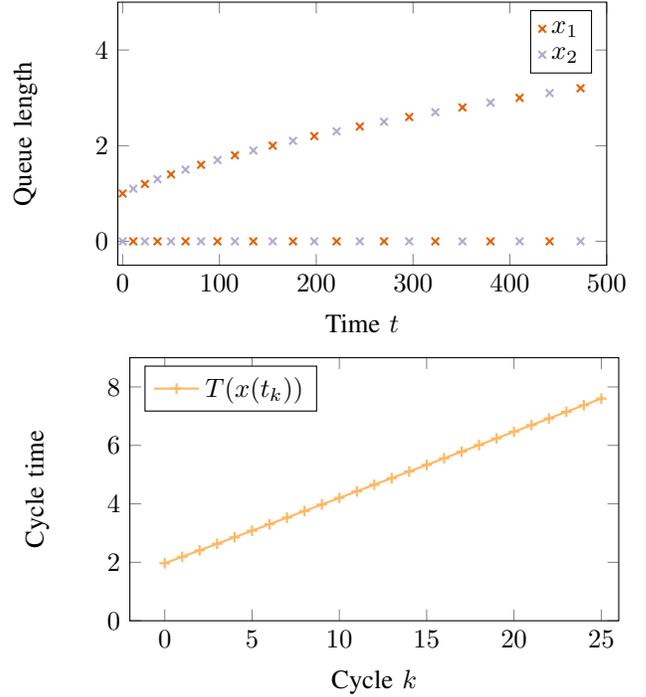
\begin{figure}
\centering
%\definecolor{mycolor1}{rgb}{0.00000,0.44700,0.74100}%
%\definecolor{mycolor2}{rgb}{0.85000,0.32500,0.09800}%
%

% lambda = [0.4; 0.3];
% kappa = 0.1;
% K = 0.1;
% x(1,1) = 6;
% x(2,1) = 4;

\begin{tikzpicture}
\begin{axis}[%
width=6.5cm,
height=3.5cm,
scale only axis,
xmin=-5,
xmax=500,
ymin=-0.5,
ymax=5,
axis background/.style={fill=white},
ylabel={Queue length},
xlabel={Time $t$},
legend pos=north east
]
\addplot [color=mycolor1, mark=x, only marks, thick]
  table[row sep=crcr]{%
0	1\\
11	0\\
23	1.2\\
36	0\\
50	1.4\\
65	0\\
81	1.6\\
98	0\\
116	1.8\\
135	0\\
155	2\\
176	0\\
198	2.2\\
221	0\\
245	2.4\\
270	0\\
296	2.6\\
323	0\\
351	2.8\\
380	0\\
410	3\\
441	0\\
473	3.2\\
506	0\\
540	3.4\\
575	0\\
611	3.6\\
648	0\\
686	3.8\\
725	0\\
765	4\\
};
\addlegendentry{$x_1$}

\addplot [color=mycolor3, mark=x, only marks, thick]
  table[row sep=crcr]{%
0	0\\
11	1.1\\
23	0\\
36	1.3\\
50	0\\
65	1.5\\
81	0\\
98	1.7\\
116	0\\
135	1.9\\
155	0\\
176	2.1\\
198	0\\
221	2.3\\
245	0\\
270	2.5\\
296	0\\
323	2.7\\
351	0\\
380	2.9\\
410	0\\
441	3.1\\
473	0\\
506	3.3\\
540	0\\
575	3.5\\
611	0\\
648	3.7\\
686	0\\
725	3.9\\
765	0\\
};
\addlegendentry{$x_2$}
\end{axis}
\end{tikzpicture}

\medskip
\begin{tikzpicture}
\begin{axis}[%
width=6.5cm,
height=3.5cm,
scale only axis,
xmin=-2,
xmax=26,
ymin=0,
ymax=9,
axis background/.style={fill=white},
ylabel={Cycle time},
xlabel={Cycle $k$},
legend pos=north west
]
\addplot [color=mycolor2, mark=+, thick]
  table[row sep=crcr]{%
0	1.96748538590988\\
1	2.18864042262606\\
2	2.41055592663846\\
3	2.6331145241309\\
4	2.85622408999971\\
5	3.07981098556073\\
6	3.30381541947856\\
7	3.52818817877235\\
8	3.75288827118614\\
9	3.97788118999694\\
10	4.20313761387689\\
11	4.42863241711839\\
12	4.65434390533815\\
13	4.88025321768828\\
14	5.10634385385028\\
15	5.33260129579977\\
16	5.55901270242869\\
17	5.78556666080505\\
18	6.01225298191313\\
19	6.23906253165705\\
20	6.46598709006465\\
21	6.69301923322637\\
22	6.92015223370069\\
23	7.14737997602479\\
24	7.37469688466176\\
25	7.60209786225012\\
%26	7.82957823643693\\
%27	8.05713371390134\\
%28	8.28476034043202\\
%29	8.51245446612666\\
%30	8.74021271494427\\
%31	0\\
};
\addlegendentry{$T(x(t_k))$}

\end{axis}
\end{tikzpicture}%
\caption{How the traffic volumes evolve in time together with the cycle times for the system in Example~\ref{ex:unstableunboundedcycle}. We can observe that the cycle length increases for each cycle.}
\label{fig:unstableunboundedcycle}
\end{figure}
\medskip 
Imposing an upper bound on the cycle length, and hence a lower bound on $w$ will then shrink the throughput region. An upper bound of the cycle length may occurs naturally, due to the fact that the sensors cover a limited area and hence the measurements will saturate. However, we will later observe in the simulations that $\bar{w} > 0$ may improve the performance of the controller when it is simulated in a realistic scenario, even when saturation of the queue length measurements is possible.

\subsection{GPA with Shorted Cycles}\label{sec:GPAshorted}
One possible drawback of the controller in Section~\ref{sec:GPAfull} is that it has to activate all the clearance phases in one cycle. This property implies that if the junction is empty when the signal program is computed, it will take $n_{p_j} T_w$ seconds until a new signal program is computed. Motivated by this, we also present a version of the GPA where only the clearance phases get activated if their corresponding phases have been activated. If we let $n_{p_j}'$ denote the number of phases that will be activated during the upcoming cycle, the total cycle time is given by
$$T_\text{cyc} = \frac{n_{p_j}' T_w}{w} \, .$$
How to compute the signal program in this case, is shown in Algorithm~\ref{algo:gpashorted}.
\begin{figure}[!t]
\removelatexerror
\begin{algorithm}[H]
\caption{GPA with Shorted Cycles}\label{algo:gpashorted}
\DontPrintSemicolon
\KwData{Current time $t$, local queue lengths $x^{(j)}(t)$, phase matrix $P^{(j)}$, clearance time $T_w$, tuning parameters $\kappa, \bar w$}
\KwResult{Signal program $\mc T^{(j)}$}
%\Begin{
   $\mc T^{(j)} \leftarrow \emptyset$ \;
   $n_{p_j} \leftarrow $ Number of rows in $P^{(j)}$ \;
   $(\nu, w)$ $\leftarrow$ Solution to~\eqref{eq:gpa} given $x^{(j)}(t), P^{(j)}, \kappa, \bar w$\;
   \Comment*[l]{Compute the number of phases to be activated} 
    $n_{p_j}' \leftarrow 0$ \;
   \For{$i\leftarrow 1$ \KwTo $n_{p_j}$}{
   	\If{$\nu_i > 0$}{
		$n_{p_j} ' \leftarrow n_{p_j}' + 1$ \;
	}
    }
    
   \uIf{$n_{p_j} ' > 0$}{
   	   	\Comment*[l]{If vehicles are present on some phases, activate those}

       $T_\text{cyc} \leftarrow n'_{p_j} \cdot T_w / w$ \;
       $t_\text{end} \leftarrow t$ \;
      \For{$i\leftarrow 1$ \KwTo $n_p$}{
      	  \If{$\nu_i > 0$} {
         	$t_\text{end} \leftarrow t_\text{end} + \nu_i \cdot T_\text{cyc}$ \;
	 \Comment*[l]{Add phase $p_i$} 
    	$\mc T^{(j)} \leftarrow \mc T^{(j)} + (p_i, t_\text{end})$ \;
    	$t_\text{end} \leftarrow t_\text{end} + T_w$ \;
	\Comment*[l]{Add clearance phase $p'_i$ }
             $\mc T^{(j)} \leftarrow \mc T^{(j)} +  (p'_i, t_\text{end})$  
       }}
   }
   \Else{
   	\Comment*[l]{If no vehicles are present, hold a clearance phase for one time unit}
   	 $\mc T^{(j)} \leftarrow (p'_1, t+1)$ 
   }
%}
\end{algorithm}
\end{figure}
\subsection{MaxPressure}
As mentioned in the introduction, the MaxPressure controller is another throughput optimal feedback controller for traffic signals. The controller computes the difference between the queue lengths and their downstream queue lengths in each phase, to determine each phase's pressure. It then activates the phase with the most pressure for a fixed time interval. To compute the pressure, the controller needs information about where the outflow from every queue will proceed. To model this, we introduce the routing matrix $R \in \R_+^{\mc E \times \mc E}$, whose elements $R_{ij}$ tells the fraction of vehicles that will proceed from lane $i$ in the current junction to lane $j$ in a downstream junction.

With the knowledge of the routing matrix and under the assumption that the flow rates are the same for all phases, the pressure, $w_i$, for each phase $p_i \in \mc P^{j}$ can then be computed as
$$w_i = \sum_{l \in p_i} \biggl( x_l(t) - \sum_k R_{lk} x_k(t) \biggr) \, .$$
The phase that should be activated is then any phase in the set $ \argmax_i w_i \,.$

Apart from the routing matrix, the MaxPressure controller has one tuning parameter, the phase duration $d > 0$. That parameter tells how long a phase should be activated, and hence how long it should take until the pressures are resampled, and a new phase activation decision is made.

How to compute the signal program with the MaxPressure controller is shown in Algorithm~\ref{algo:maxpressure}.
\begin{figure}[!t]
 \removelatexerror
  \begin{algorithm}[H] \DontPrintSemicolon

   \caption{MaxPressure}\label{algo:maxpressure}
   \KwData{Current time $t$, local queue lengths $x(t)$, phase matrix $P^{(j)}$, routing matrix $R$, phase duration $d$}
   \KwResult{Signal program $\mc T^{(j)}$}  
   $\mc T^{(j)} \leftarrow \emptyset$ \;
   $n_{p_j} \leftarrow $ Number of rows in $P^{(j)}$ \;
   \For{$i\leftarrow 1$ \KwTo $n_{p_j} $}{
   	\For{$l \in \mc L^{(j)}$} {
		\If{$l \in p_i^{(j)}$} {
		 	$w_i \leftarrow w_i + x_l(t) - \sum_{k} R_{lk} x_k(t)$
		}
	}
   }
   $i \leftarrow \argmax_i w_i$ \;
   \Comment*[l]{Add phase $p_i$} 
   $\mc T^{(j)} \leftarrow \mc T^{(j)} + (p_i, t + d)$ \;  
    \Comment*[l]{Add clearance phase $p'_i$}    
   $\mc T^{(j)} \leftarrow \mc T^{(j)} + (p'_i, t+ d + T_w)$ \; 
  \end{algorithm}
\end{figure}

\section{Comparison Between GPA and MaxPressure} \label{sec:comparision}
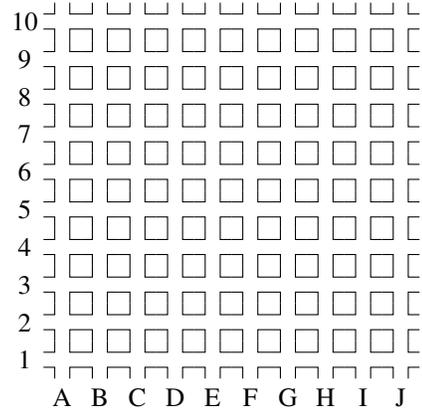
\begin{figure}
\centering
\begin{tikzpicture}[scale=0.5]
\foreach \i in {1,...,10}
{
	\foreach \j in {1,...,10} {
        \pgfmathtruncatemacro{\label}{\i \j};
		\draw (-0.5 + \i,\j+0.2) -- (\i-0.2, \j+0.2) -- (\i-0.2, \j+0.5);
		\draw (0.2+ \i,\j+0.5) -- (\i+0.2, \j+0.2) -- (\i+0.5, \j+0.2);
		\draw (0.2+ \i,\j-0.5) -- (\i+0.2, \j-0.2) -- (\i+0.5, \j-0.2);
		\draw (-0.5 + \i,\j-0.2) -- (\i-0.2, \j-0.2) -- (\i-0.2, \j-0.5);
		}
}
\foreach \i in {1,...,10}
{
	\node (Y\i) at (0,\i) {\i};
}
\foreach[count=\i] \j in {A,...,J}
{
	\node (X\j) at (\i, 0) {\j};
}
\end{tikzpicture}
\caption{The Manhattan-like network used in the comparison between GPA and MaxPressure. }
\label{fig:network}
\end{figure}

\begin{figure}
\centering
\begin{tabular}{cc}

\begin{tikzpicture}[scale=0.5]
\draw (-3, 0) -- (-1,0);
\draw (1,0) -- (3,0);
\draw (0,1) -- (0,3);
\draw (0,-1) -- (0,-3);

\draw[thick]  (1,-3)  -- (1,-1) -- (1.5, -0.5) -- (3, -0.5) ;
\draw[thick] (-1,-1) -- (-0.5, -1.5) -- (-0.5,-3) ;

\draw[thick] (-3, -1) -- (-1,-1);

\draw[thick] (-3, 0.5) -- (-1.5, 0.5)--(-1,1);
\draw[thick] (-1, 1) -- (-1, 3);

\draw[thick] (0.5, 3) -- (0.5,1.5) -- (1,1) -- (3,1);
\draw[dashed] (1, 0.5) -- (3,0.5);

\draw[->] (0.75, -1.5) -- (0.75, -1.1);
\draw[->] (0.75, -1.3) -- (0.95, -1.3);
\draw[->] (0.25, -1.5) -- (0.25, -1.3) -- (0.05,-1.3);

\draw[->] (-1.5, -0.75) -- ( -1.1, -0.75);
\draw[->] (-1.3, -0.75) -- ( -1.3, -0.95);
\draw[->] (-1.5, -0.25) -- (-1.3, -0.25) -- (-1.3, -0.05);

\draw[->] (-0.75, 1.5) -- (-0.75, 1.1);
\draw[->] (-0.75, 1.3) -- (-0.95, 1.3);
\draw[->] (-0.25, 1.5) -- (-0.25, 1.3) -- (-0.05,1.3);

\draw[->] (1.5, 0.75) -- ( 1.1, 0.75);
\draw[->] (1.3, 0.75) -- ( 1.3, 0.95);
\draw[->] (1.5, 0.25) -- (1.3, 0.25) -- (1.3, 0.05);

\draw[mycolor1, thick] (-0.75, 1) edge[bend right=10, ->] (-0.25,-1);
\draw[mycolor1, thick] (-0.75, 1) edge[bend left=30, ->] (-1,0.25);
\draw[mycolor1, thick] (0.75, -1) edge[bend left=10, ->] (0.25,1);
\draw[mycolor1, thick] (0.75, -1) edge[bend left=30, ->] (1,-0.25);

\draw[mycolor2, thick] (-0.25, 1) edge[bend right=10, ->] (1,-0.25);
\draw[mycolor2, thick] (0.25, -1) edge[bend right=10, ->] (-1,0.25);

\draw[mycolor3,thick] (1, 0.75) edge[bend left=30, ->] (0.25,1);
\draw[mycolor3,thick] (1, 0.75) edge[bend right=10, ->] (-1,0.25);
\draw[mycolor3,thick] (-1, -0.75) edge[bend left=30, ->] (-0.25, -1);
\draw[mycolor3,thick] (-1, -0.75) edge[bend right=10, ->] (1, -0.25);

\draw[mycolor4, thick] (1,0.25) edge[bend right=10, ->] (-0.25, -1);
\draw[mycolor4, thick] (-1,-0.25) edge[bend right=10, ->] (0.25, 1);

\draw[dashed] (0.5, -1) -- (0.5,-3);
\draw[dashed] (-3, -0.5) -- (-1, -0.5);
\draw[dashed] (-3, -0.5) -- (-1, -0.5);
\draw[dashed] (-0.5,1) -- (-0.5,3);

\end{tikzpicture} 
&
\begin{tikzpicture}[scale=0.5]

\draw (-3, 0) -- (-1,0);
\draw (2,0) -- (4,0);
\draw (0.5,1) -- (0.5,3);
\draw (0.5,-1) -- (0.5,-3);

\draw[thick]  (2,-3)  -- (2,-1) -- (2.5, -0.5) -- (4, -0.5) ;

\draw[thick] (-1,-1) -- (-0.5, -1.5) -- (-0.5,-3) ;

\draw[thick] (-3, -1) -- (-1,-1);

\draw[thick] (-3, 0.5) -- (-1.5, 0.5)--(-1,1);
\draw[thick] (-1, 1) -- (-1, 3);

\draw[thick] (1.5, 3) -- (1.5,1.5) -- (2,1) -- (4,1);

\draw[->] (1.25, -1.5) -- (1.25, -1.1);
\draw[->] (1.75, -1.5) -- (1.75, -1.1);
\draw[->] (1.75, -1.3) -- (1.95, -1.3);
\draw[->] (0.75, -1.5) -- (0.75, -1.3) -- (0.55,-1.3);

\draw[->] (-1.5, -0.75) -- ( -1.1, -0.75);
\draw[->] (-1.3, -0.75) -- ( -1.3, -0.95);
\draw[->] (-1.5, -0.25) -- (-1.3, -0.25) -- (-1.3, -0.05);

\draw[->] (-0.75, 1.5) -- (-0.75, 1.1);
\draw[->] (-0.75, 1.3) -- (-0.95, 1.3);
\draw[->] (-0.25, 1.5) -- (-0.25, 1.1);

\draw[->] (0.25, 1.5) -- (0.25, 1.3) -- (0.45,1.3);

\draw[->] (2.5, 0.75) -- ( 2.1, 0.75);
\draw[->] (2.3, 0.75) -- ( 2.3, 0.95);
\draw[->] (2.5, 0.25) -- (2.3, 0.25) -- (2.3, 0.05);

\draw[mycolor1, thick] (-0.75, 1) edge[bend right=10, ->] (-0.25,-1);
\draw[mycolor1, thick] (-0.25, 1) edge[bend right=10, ->] (0.25,-1);
\draw[mycolor1, thick] (-0.75, 1) edge[bend left=30, ->] (-1,0.25);

\draw[mycolor1, thick] (1.75, -1) edge[bend left=10, ->] (1.25,1);
\draw[mycolor1, thick] (1.75, -1) edge[bend left=30, ->] (2,-0.25);
\draw[mycolor1, thick] (1.25, -1) edge[bend left=10, ->] (.75,1);

\draw[mycolor2, thick] (0.25, 1) edge[bend right=10, ->] (2,-0.25);
\draw[mycolor2, thick] (0.75, -1) edge[bend right=10, ->] (-1,0.25);

\draw[mycolor3,thick] (2, 0.75) edge[bend left=30, ->] (1.25,1);
\draw[mycolor3,thick] (2, 0.75) edge[bend right=10, ->] (-1,0.25);

\draw[mycolor3,thick] (-1, -0.75) edge[bend left=30, ->] (-0.25, -1);
\draw[mycolor3,thick] (-1, -0.75) edge[bend right=10, ->] (2, -0.25);

\draw[mycolor4, thick] (2,0.25) edge[bend right=10, ->] (0.25, -1);
\draw[mycolor4, thick] (-1,-0.25) edge[bend right=10, ->] (0.75, 1);

\draw[dashed] (0, -1.5) -- (0,-3);
\draw[dashed] (1 ,-1) -- (1,-3);
\draw[dashed] (1.5,-1) -- (1.5,-3);

\draw[dashed] (-3, -0.5) -- (-1, -0.5);
\draw[dashed] (-0.5,1) -- (-0.5,3);
\draw[dashed] (-0,1) -- (-0,3);
\draw[dashed] (2, 0.5) -- (4,0.5);

\draw[dashed] (1,1.5) --(1,3);

\end{tikzpicture} \\
2 by 2 junction & 2 by 3 junction
\\ 
 & \\
\begin{tikzpicture}[scale=0.5]
\begin{scope}[rotate=-90]
\draw (-3, 0) -- (-1,0);
\draw (2,0) -- (4,0);
\draw (0.5,1) -- (0.5,3);
\draw (0.5,-1) -- (0.5,-3);

\draw[thick]  (2,-3)  -- (2,-1) -- (2.5, -0.5) -- (4, -0.5) ;

\draw[thick] (-1,-1) -- (-0.5, -1.5) -- (-0.5,-3) ;

\draw[thick] (-3, -1) -- (-1,-1);

\draw[thick] (-3, 0.5) -- (-1.5, 0.5)--(-1,1);
\draw[thick] (-1, 1) -- (-1, 3);

\draw[thick] (1.5, 3) -- (1.5,1.5) -- (2,1) -- (4,1);

\draw[->] (1.25, -1.5) -- (1.25, -1.1);
\draw[->] (1.75, -1.5) -- (1.75, -1.1);
\draw[->] (1.75, -1.3) -- (1.95, -1.3);
\draw[->] (0.75, -1.5) -- (0.75, -1.3) -- (0.55,-1.3);

\draw[->] (-1.5, -0.75) -- ( -1.1, -0.75);
\draw[->] (-1.3, -0.75) -- ( -1.3, -0.95);
\draw[->] (-1.5, -0.25) -- (-1.3, -0.25) -- (-1.3, -0.05);

\draw[->] (-0.75, 1.5) -- (-0.75, 1.1);
\draw[->] (-0.75, 1.3) -- (-0.95, 1.3);
\draw[->] (-0.25, 1.5) -- (-0.25, 1.1);

\draw[->] (0.25, 1.5) -- (0.25, 1.3) -- (0.45,1.3);

\draw[->] (2.5, 0.75) -- ( 2.1, 0.75);
\draw[->] (2.3, 0.75) -- ( 2.3, 0.95);
\draw[->] (2.5, 0.25) -- (2.3, 0.25) -- (2.3, 0.05);

\draw[mycolor1, thick] (-0.75, 1) edge[bend right=10, ->] (-0.25,-1);
\draw[mycolor1, thick] (-0.25, 1) edge[bend right=10, ->] (0.25,-1);
\draw[mycolor1, thick] (-0.75, 1) edge[bend left=30, ->] (-1,0.25);

\draw[mycolor1, thick] (1.75, -1) edge[bend left=10, ->] (1.25,1);
\draw[mycolor1, thick] (1.75, -1) edge[bend left=30, ->] (2,-0.25);
\draw[mycolor1, thick] (1.25, -1) edge[bend left=10, ->] (.75,1);

\draw[mycolor2, thick] (0.25, 1) edge[bend right=10, ->] (2,-0.25);
\draw[mycolor2, thick] (0.75, -1) edge[bend right=10, ->] (-1,0.25);

\draw[mycolor3,thick] (2, 0.75) edge[bend left=30, ->] (1.25,1);
\draw[mycolor3,thick] (2, 0.75) edge[bend right=10, ->] (-1,0.25);

\draw[mycolor3,thick] (-1, -0.75) edge[bend left=30, ->] (-0.25, -1);
\draw[mycolor3,thick] (-1, -0.75) edge[bend right=10, ->] (2, -0.25);

\draw[mycolor4, thick] (2,0.25) edge[bend right=10, ->] (0.25, -1);
\draw[mycolor4, thick] (-1,-0.25) edge[bend right=10, ->] (0.75, 1);
\draw[dashed] (0, -1.5) -- (0,-3);
\draw[dashed] (1 ,-1) -- (1,-3);
\draw[dashed] (1.5,-1) -- (1.5,-3);
\draw[dashed] (-3, -0.5) -- (-1, -0.5);
\draw[dashed] (-0.5,1) -- (-0.5,3);
\draw[dashed] (-0,1) -- (-0,3);
\draw[dashed] (2, 0.5) -- (4,0.5);
\draw[dashed] (1,1.5) --(1,3);
\end{scope}
\end{tikzpicture}
&
\begin{tikzpicture}[scale=0.5]

\draw (-3, 0) -- (-1,0);
\draw (2,0) -- (4,0);
\draw (0.5,1.5) -- (0.5,3.5);
\draw (0.5,-1.5) -- (0.5,-3.5);

\draw[thick]  (2,-3.5)  -- (2,-1.5) -- (2.5, -1) -- (4, -1) ;

\draw[thick] (-1,-1.5) -- (-0.5, -2) -- (-0.5,-3.5) ;

\draw[thick] (-3, -1.5) -- (-1,-1.5);

\draw[thick] (-3, 1) -- (-1.5, 1)--(-1,1.5);
\draw[thick] (-1, 1.5) -- (-1, 3.5);

\draw[thick] (1.5, 3.5) -- (1.5,2) -- (2,1.5) -- (4,1.5);

\draw[->] (1.25, -2) -- (1.25, -1.6);
\draw[->] (1.75, -2) -- (1.75, -1.6);
\draw[->] (1.75, -1.8) -- (1.95, -1.8);
\draw[->] (0.75, -2) -- (0.75, -1.8) -- (0.55,-1.8);

\draw[->] (-1.5, -1.25) -- ( -1.1, -1.25);
\draw[->] (-1.3, -1.25) -- ( -1.3, -1.45);
\draw[->] (-1.5, -0.25) -- (-1.3, -0.25) -- (-1.3, -0.05);
\draw[->] (-1.5, -0.75) -- (-1.1, -0.75);

\draw[->] (-0.75, 2) -- (-0.75, 1.6);
\draw[->] (-0.75, 1.8) -- (-0.95, 1.8);
\draw[->] (-0.25, 2) -- (-0.25, 1.6);

\draw[->] (0.25, 2) -- (0.25, 1.8) -- (0.45,1.8);

\draw[->] (2.5, 1.25) -- ( 2.1, 1.25);
\draw[->] (2.3, 1.25) -- ( 2.3, 1.45);
\draw[->] (2.5, 0.75) -- ( 2.1, 0.75);

\draw[->] (2.5, 0.25) -- (2.3, 0.25) -- (2.3, 0.05);

\draw[mycolor1, thick] (-0.75, 1.5) edge[bend right=10, ->] (-0.25,-1.5);
\draw[mycolor1, thick] (-0.25, 1.5) edge[bend right=10, ->] (0.25,-1.5);
\draw[mycolor1, thick] (-0.75, 1.5) edge[bend left=30, ->] (-1,0.75);

\draw[mycolor1, thick] (1.75, -1.5) edge[bend left=10, ->] (1.25,1.5);
\draw[mycolor1, thick] (1.75, -1.5) edge[bend left=30, ->] (2,-0.75);
\draw[mycolor1, thick] (1.25, -1.5) edge[bend left=10, ->] (.75,1.5);

\draw[mycolor2, thick] (0.25, 1.5) edge[bend right=10, ->] (2,-0.25);
\draw[mycolor2, thick] (0.75, -1.5) edge[bend right=10, ->] (-1,0.25);

\draw[mycolor3,thick] (2, 1.25) edge[bend left=30, ->] (1.25,1.5);
\draw[mycolor3,thick] (2, 1.25) edge[bend right=10, ->] (-1,0.75);
\draw[mycolor3,thick] (2, 0.75) edge[bend right=10, ->] (-1,0.25);

\draw[mycolor3,thick] (-1, -1.25) edge[bend left=30, ->] (-0.25, -1.5);
\draw[mycolor3,thick] (-1, -1.25) edge[bend right=10, ->] (2, -0.75);
\draw[mycolor3,thick] (-1, -0.75) edge[bend right=10, ->] (2, -0.25);

\draw[mycolor4, thick] (2,0.25) edge[bend right=10, ->] (0.25, -1.5);
\draw[mycolor4, thick] (-1,-0.25) edge[bend right=10, ->] (0.75, 1.5);

\draw[dashed] (0, -2) -- (0,-3.5);
\draw[dashed] (1 ,-1.5) -- (1,-3.5);
\draw[dashed] (1.5,-1.5) -- (1.5,-3.5);

\draw[dashed] (-3, -1) -- (-1, -1);
\draw[dashed] (-3, -0.5) -- (-1, -0.5);
\draw[dashed] (-3, 0.5) -- (-1.5, 0.5);

\draw[dashed] (-0.5,1.5) -- (-0.5,3.5);
\draw[dashed] (-0,1.5) -- (-0,3.5);

\draw[dashed] (2, 1) -- (4,1);
\draw[dashed] (2, 0.5) -- (4,0.5);

\draw[dashed] (1,2) --(1,3.5);
\draw[dashed] (2.5, -0.5) -- (4, -0.5);

\end{tikzpicture} \\

3 by 2 junction & 3 by 3 junction
\end{tabular}
\caption{The four different types of junctions present in the Manhattan grid, together with theirs phases.} \label{fig:junction}

\end{figure}
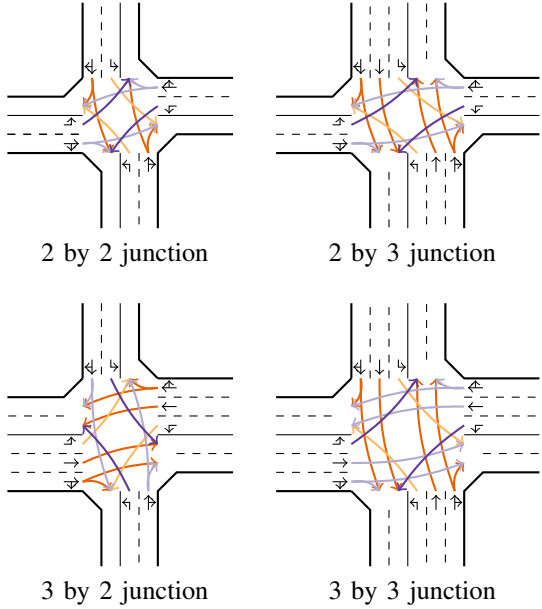

\subsection{Simulation setting}

To compare the proposed controller and the MaxPressure controller, we simulate both controllers on an artificial Manhattan-like grid with artificial demand.
The simulator we are using is open source micro simulator SUMO~\cite{SUMO2012}, which is a simulator that simulates every single vehicle's behavior in the traffic network.
A schematic drawing of the network is shown in Fig.~\ref{fig:network}. In a setting like this, we can elaborate with the tuning ratios, and provide the MaxPressure controller both correct and incorrect turning ratios. This allows us to investigate the robustness properties of both the controllers.

The Manhattan grid in Fig.~\ref{fig:network}  has ten bidirectional north to south streets (indexed A to J) and ten bidirectional east to west streets (indexed 1 to 10). All streets with an odd number or indexed by letter A, C, E, G or I consist of one lane in each direction, while the others consist of two lanes in each direction. The speed limit on each lane is 50 km/h. The distance between each junction is three hundred meters. Fifty meters before each junction, every street has an additional lane, reserved for vehicles that want to turn left. Due to the varying number of lanes, four different junction topologies exist, all shown in Fig.~\ref{fig:junction}, together with the set of possible phases. Each junction is equipped with sensors on the incoming lanes that can measure the number of vehicles queuing up to fifty meters from the junction. The sensors measure the queue lengths by the number of stopped vehicles.

Since the scenario is artificial, we can generate demand with prescribed turning ratios and hence let the MaxPressure controller to run in an ideal setting. For the demand generation, we assume that at each junction a vehicle will with probability $0.2$ will turn left, with probability $0.6$ go straight and with probability $0.2$ turn right. We do assume that all vehicles depart from lanes connected to the boundary of the network, and all vehicles will also end their trips when they have reached the boundary of the network. In other words, no vehicles will depart or arrive inside the grid. We will study the controllers' performance for three different demands, where the demand determined by the probability that a vehicle will depart from each boundary lane each second. We denote this probability $\delta$, where the probabilities for the three different demands are $\delta = 0.05$, $\delta = 0.1$ and $\delta = 0.15$. We generate vehicles for $3600$ seconds and then simulate until all vehicles have left the network.

We also compare the results for the GPA controller and the MaxPressure controller with a standard fixed time (FT) controller and a proportional fair (PF) controller, i.e., the GPA controller with full clearance cycles, but with $\kappa =0$ and a prescribed fixed cycle length. For the fixed time controller, the phases which contain a straight movement are activated for $30$ seconds and phases only containing left or right turn movements are activated for $15$ seconds. The clearance time for each phase is still set to $5$ seconds. This means that the cycle lengths for each of the four types of junctions will be $110$ seconds. This is also the fixed cycle time we are using for the proportional fairness controller.

\subsection{GPA Results}
\begin{figure}
\centering
\begin{tikzpicture}
\begin{axis}[ymode=log, width=8cm,  height=6cm,  ylabel={Total Queue Length [m] }, xlabel={Time [s]}, xmax=6000, legend style={at={(0.5,-0.25)},anchor=north}]
\addplot[mark=none, color=mycolor1, thick] table [x index=0, y index=1]{plotdata/bpvspf/queue_pf2_k10_l0.05.csv};
\addplot[mark=none, color=mycolor2, thick] table [x index=0, y index=1]{plotdata/bpvspf/queue_pf2_k10_l0.10.csv};
\addplot[mark=none, color=mycolor3, thick] table [x index=0, y index=1]{plotdata/bpvspf/queue_pf2_k10_l0.15.csv};
\addplot[mark=none, color=mycolor1, dashed, thick] table [x index=0, y index=1]{plotdata/bpvspf/queue_pf2_k5_l0.05.csv};
\addplot[mark=none, color=mycolor2, dashed, thick] table [x index=0, y index=1]{plotdata/bpvspf/queue_pf2_k5_l0.10.csv};
\addplot[mark=none, color=mycolor3, dashed, thick] table [x index=0, y index=1]{plotdata/bpvspf/queue_pf2_k5_l0.15.csv};
\legend{GPA $\kappa=10 \, \delta = 0.05$, GPA $\kappa=10 \, \delta = 0.10$, GPA $\kappa=10 \, \delta = 0.15$ , GPA $\kappa=5 \, \delta = 0.05$, GPA $\kappa=5 \, \delta = 0.10$, GPA $\kappa=5 \, \delta = 0.15$ }
\end{axis}
\end{tikzpicture}
\caption{How the queue length varies with time when the GPA with shorted cycles are used in Manhattan grid. The GPA is tested with two different values of $\kappa=5,10$ for the three demand scenarios $\delta = 0.05, 0.10, 0.15$. To improve the readability of the results, the queue-lengths are averaged over $300$ seconds intervals.}
\label{fig:gpamanhattan}
\end{figure}
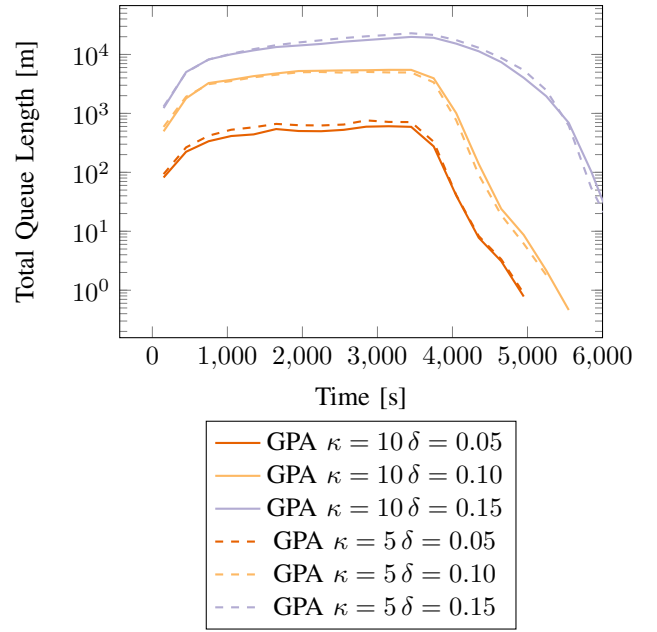

Since the phases in this scenario are all orthogonal, the expressions in~\eqref{eq:gpaorthogonal} can be used to solve the optimization problem in~\eqref{eq:gpa}. The tuning parameter $\bar{w}$ is set to $\bar{w} = 0$ for all simulations.  In Table~\ref{tab:gpamanhattan} we show how the total travel time varies for the GPA controller with shorted cycles for different values of $\kappa$. For the demand $\delta = 0.15$ and $\kappa =1$ a gridlock situation occurs, probably due to the fact that vehicles back-spills into upstream junctions. We can see that a $\kappa =10$ seems to be the best choice for $\delta = 1$ and $\delta = 0.15$, while a higher $\kappa$ slightly improves the total travel time for the lowest demand investigated. Letting $\kappa = 10$ has been shown to be reasonable for other demand scenarios in the same network setting, as observed in~\cite{nilsson2018}. How the total queue lengths varies with time for $\kappa =5$ and $\kappa = 10$ is shown in Fig.~\ref{fig:gpamanhattan}.
\begin{table}
\centering
\caption{GPA with Shorted Cycles - Manhattan Scenario}
\label{tab:gpamanhattan}
\begin{tabular}{rcc}
$\kappa$ & $\delta$ & Total Travel Time [h] \\ \hline \hline
$1$ & $0.05$ &  $1398$ \\
$5$ & $0.05$ &  \phantom{0}$715$ \\
$10$ & $0.05$ & \phantom{0}$699$   \\
$15$ & $0.05$ & \phantom{0}$696$   \\
$20$ & $0.05$ & \phantom{0}$690$   \\
$1$ & $0.10$ & $7636$ \\
$5$ & $0.10$ &  $1898$ \\
$10$ & $0.10$ &  $1992$ \\
$15$ & $0.10$ &  $2263$ \\
$20$ & $0.10$ &  $2495$ \\
$1$ & $0.15$ & $+\infty$ \\
$5$ & $0.15$ &  $5134$ \\
$10$ & $0.15$ &  $4498$ \\
$15$ & $0.15$ &  $5140$ \\
$20$ & $0.15$ &  $6050$ \\ \hline
\end{tabular}
\end{table}

\subsection{MaxPressure Results}
The MaxPressure controller decides its control action not only based on queue-lengths on the incoming lanes, but also on the downstream lanes. It is not always clear in which downstream lane a vehicle will end up in after leaving the junction. If a vehicle can choose between several lanes that are all valid for its path, the vehicle's lane choice will be determined during the simulation, and depend upon how many other vehicles that are occupying the possible lanes. Because of this, we assume that if a vehicle can choose between several lanes, it will try to join the shortest one. To exemplify how the turning ratios are estimated in those situations, assume that Moreover, assume that the overall probability  that a vehicle is turning right is $0.2$, and going straight is $0.6$. If a vehicle going straight can choose between lane $l_1$, $l_2$, but $l_2$ is also used by vehicles turning right, the probability that the vehicle going straight will queue up in lane $l_1$ is assumed to be $0.4$ and that the probability that the vehicle will queue up in lane $l_2$ is estimated to be $0.2$.

To also investigate the MaxPressure controller's robustness with respect to the routing information, we perform simulations both when the controller has the correct information about the turning probabilities, i.e., that a vehicle will turn right with probability $0.2$, continue straight with probability $0.6$ and turn left with probability $0.2$. For the simulations when the MaxPressure has the wrong turning information, the controller instead has the information that with probability $0.6$ the vehicle will turn right, with probability $0.3$ the vehicle will proceed straight and with probability $0.1$ the vehicle will turn left. In the simulations, we consider three different phase durations, $d=10$ seconds, $d=20$ seconds and $d=30$ seconds. 

\begin{figure}
\centering
\begin{tikzpicture}
\begin{axis}[width=8cm,  height=6cm,  ylabel={Total Queue Length [m] }, xlabel={Time [s]}, xmax=6000, legend pos=north west, scaled y ticks = false,
      y tick label style={/pgf/number format/fixed,
      /pgf/number format/1000 sep = \thinspace % Optional if you want to replace comma as the 1000 separator 
      },  legend style={at={(0.5,-0.25)},anchor=north}]
\addplot[mark=none, color=mycolor1, thick] table [x index=0, y index=1]{plotdata/bpvspf/queue_bp_d10_l0.15.csv};
\addplot[mark=none, color=mycolor2, thick] table [x index=0, y index=1]{plotdata/bpvspf/queue_bp_d20_l0.15.csv};
\addplot[mark=none, color=mycolor3, thick] table [x index=0, y index=1]{plotdata/bpvspf/queue_bp_d30_l0.15.csv};
\addplot[mark=none, color=mycolor1, dashed, thick] table [x index=0, y index=1]{plotdata/bpvspf/queue_bpw_d10_l0.15.csv};
\addplot[mark=none, color=mycolor2, dashed, thick] table [x index=0, y index=1]{plotdata/bpvspf/queue_bpw_d20_l0.15.csv};
\addplot[mark=none, color=mycolor3, dashed, thick] table [x index=0, y index=1]{plotdata/bpvspf/queue_bpw_d30_l0.15.csv};
 \legend{MP $d =10$, MP $d=20$, MP $d=30$}
\end{axis}
\end{tikzpicture}
\caption{The total queue length over time in the Manhattan grid with the MaxPressure (MP) controller with right turning ratios (solid) and wrong turning ratios (dashed). The demand is $\delta = 0.15$. To improve the readability of the results, the queue-lengths are averaged over $300$ seconds intervals.}
\label{fig:mpmanhattand0.15}
\end{figure}
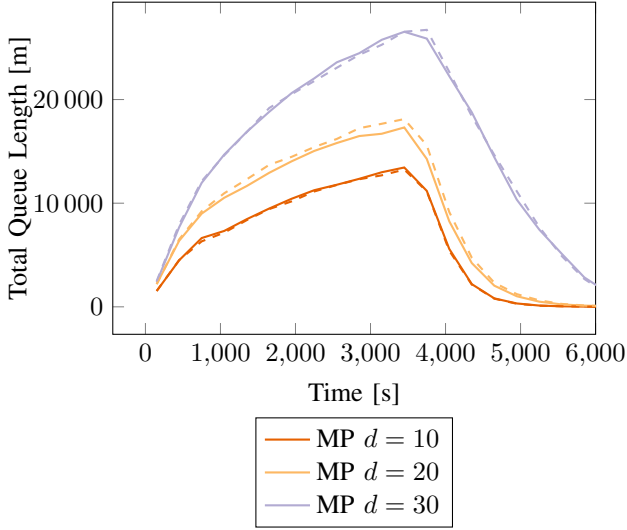

How the total queue lengths vary over time for the different demands is shown in Fig.~\ref{fig:mpmanhattand0.05}, Fig.~\ref{fig:mpmanhattand0.10}, and~Fig.~\ref{fig:mpmanhattand0.15}. The total travel time, both when the MaxPressure controller is operating with the right, and the wrong turning ratios are shown in Table~\ref{tab:mpmanhattan}. From these results, we can conclude that a shorter phase duration, i.e., $d = 10$, is the most efficient for all demands. This probably has to do with a longer phase duration the activation time is becoming larger than the time it takes to empty the measurable part of the queue. Another interesting observation is that if the MaxPressure controller has wrong information about the turning ratios, its performance does not decrease significantly.

\begin{table}
\centering
\caption{MaxPressure - Manhattan Scenario}
\label{tab:mpmanhattan}
\begin{tabular}{cccc}
 $d$ & $\delta$ & TTT correct TR [h] & TTT incorrect TR [h] \\ \hline \hline
$10$ & $0.05$ &  858 &   856\\
$20$ & $0.05$ &  1 079 & 1 102 \\
$30$ & $0.05$ &  1 172 & 1 193 \\
$10$ & $0.10$ &  1 865 & 1 864 \\
$20$ & $0.10$ &  2 254 & 2 312 \\
$30$ & $0.10$ &  2 690 & 2 718 \\
$10$ & $0.15$ &  3 511 & 3 488 \\
$20$ & $0.15$ &  3 992 & 4 102  \\
$30$ & $0.15$ &  5 579 & 5 590 \\ \hline
\end{tabular}

\end{table}

\begin{figure}
\centering
\begin{tikzpicture}
\begin{axis}[width=8cm,  height=6cm,  ylabel={Total Queue Length [m] }, xlabel={Time [s]}, xmax=5000, legend pos=north west, scaled y ticks = false,
      y tick label style={/pgf/number format/fixed,
      /pgf/number format/1000 sep = \thinspace % Optional if you want to replace comma as the 1000 separator 
      }, ,  legend style={at={(0.5,-0.25)},anchor=north}]
\addplot[mark=none, color=mycolor1, thick] table [x index=0, y index=1]{plotdata/bpvspf/queue_bp_d10_l0.05.csv};
\addplot[mark=none, color=mycolor2, thick] table [x index=0, y index=1]{plotdata/bpvspf/queue_bp_d20_l0.05.csv};
\addplot[mark=none, color=mycolor3, thick] table [x index=0, y index=1]{plotdata/bpvspf/queue_bp_d30_l0.05.csv};
\addplot[mark=none, color=mycolor1, dashed, thick] table [x index=0, y index=1]{plotdata/bpvspf/queue_bpw_d10_l0.05.csv};
\addplot[mark=none, color=mycolor2, dashed, thick] table [x index=0, y index=1]{plotdata/bpvspf/queue_bpw_d20_l0.05.csv};
\addplot[mark=none, color=mycolor3, dashed, thick] table [x index=0, y index=1]{plotdata/bpvspf/queue_bpw_d30_l0.05.csv};
 \legend{MP $d =10$, MP $d=20$, MP $d=30$}
\end{axis}
\end{tikzpicture}
\caption{The total queue length over time in the Manhattan grid with the MaxPressure (MP) controller with right turning ratios (solid) and wrong turning ratios (dashed). The demand is $\delta = 0.05$. To improve the readability of the results, the queue-lengths are averaged over $300$ seconds intervals.}
\label{fig:mpmanhattand0.05}
\end{figure}

\begin{figure}
\centering
\begin{tikzpicture}
\begin{axis}[width=8cm,  height=6cm,  ylabel={Total Queue Length [m] }, xlabel={Time [s]}, xmax=5000, legend pos=north west, scaled y ticks = false,
      y tick label style={/pgf/number format/fixed,
      /pgf/number format/1000 sep = \thinspace % Optional if you want to replace comma as the 1000 separator 
      },  legend style={at={(0.5,-0.25)},anchor=north}]
\addplot[mark=none, color=mycolor1, thick] table [x index=0, y index=1]{plotdata/bpvspf/queue_bp_d10_l0.10.csv};
\addplot[mark=none, color=mycolor2, thick] table [x index=0, y index=1]{plotdata/bpvspf/queue_bp_d20_l0.10.csv};
\addplot[mark=none, color=mycolor3, thick] table [x index=0, y index=1]{plotdata/bpvspf/queue_bp_d30_l0.10.csv};
\addplot[mark=none, color=mycolor1, dashed, thick] table [x index=0, y index=1]{plotdata/bpvspf/queue_bpw_d10_l0.10.csv};
\addplot[mark=none, color=mycolor2, dashed, thick] table [x index=0, y index=1]{plotdata/bpvspf/queue_bpw_d20_l0.10.csv};
\addplot[mark=none, color=mycolor3, dashed, thick] table [x index=0, y index=1]{plotdata/bpvspf/queue_bpw_d30_l0.10.csv};
 \legend{MP $d =10$, MP $d=20$, MP $d=30$}
\end{axis}
\end{tikzpicture}
\caption{The total queue length over time in the Manhattan grid with the MaxPressure (MP) controller with right turning ratios (solid) and wrong turning ratios (dashed). The demand is $\delta = 0.10$. To improve the readability of the results, the queue-lengths are averaged over $300$ seconds intervals.}
\label{fig:mpmanhattand0.10}
\end{figure}

\subsection{Summery of the Comparison}
\begin{figure}
\centering
\begin{tikzpicture}
\begin{axis}[ymode=log, width=8cm,  height=6cm,  ylabel={Total Queue Length [m] }, xlabel={Time [s]}, xmax=6500, legend pos=north west, scaled y ticks = false,
      y tick label style={/pgf/number format/fixed,
      /pgf/number format/1000 sep = \thinspace % Optional if you want to replace comma as the 1000 separator 
      }, , legend style={at={(0.5,-0.25)},anchor=north},  legend columns=2]

\addplot[mark=none, color=mycolor1, thick] table [x index=0, y index=1]{plotdata/bpvspf/queue_pf2_k5_l0.15.csv};
\addplot[mark=none, color=mycolor2, thick] table [x index=0, y index=1]{plotdata/bpvspf/queue_pf2_k10_l0.15.csv};
\addplot[mark=none, color=mycolor3, thick] table [x index=0, y index=1]{plotdata/bpvspf/queue_bp_d10_l0.15.csv};
\addplot[mark=none, color=mycolor4, thick] table [x index=0, y index=1]{plotdata/bpvspf/queue_fixed_l0.15.csv};
%\addplot[mark=none, color=black] table [x index=0, y index=1]{\plotdatadir/bpvspf/queue_pf_fixed_l0.15.csv};
\legend{GPA $\kappa =5$, GPA $\kappa =10$, MP $d = 10$, Fixed Time}
\end{axis}
\end{tikzpicture}
\caption{A comparison between different control strategies for the Manhattan grid with the demand $\delta = 0.15$. Since the proportional fair controller (PF) creates a gridlock, it is not included in the comparison. To improve the readability of the results, the queue-lengths are averaged over $300$ seconds intervals.}
\label{fig:comparisonl0.15}
\end{figure}
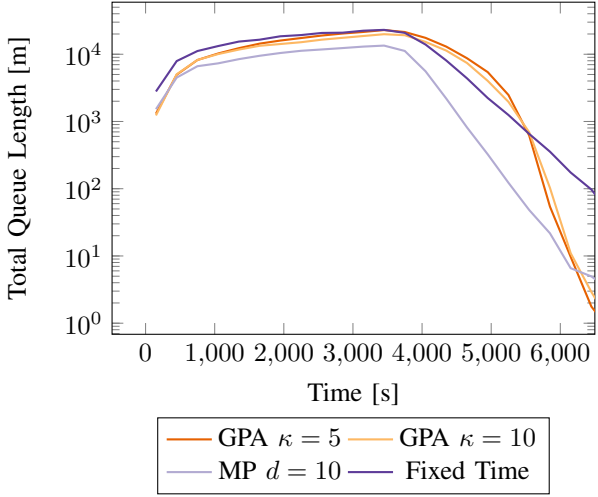

To better observe the difference between the GPA and MaxPressure, we have plotted the total queue length with the GPA controller with $\kappa = 5$ and $\kappa = 10$, and the best MaxPressure configuration with $d = 10$. The results are shown in Fig.~\ref{fig:comparisonl0.05}, Fig.~\ref{fig:comparisonl0.10} and Fig.~\ref{fig:comparisonl0.15}. In the figures we have also included for reference the total queue lengths for the fixed time controller and the proportional fair controller. The total travel travel times for those controllers are given in Table~\ref{tab:fixedmanhattan}. When the demand is $\delta = 0.15$, a gridlock situation occurs with the proportional fair controller, just as happened with the GPA controller with $\kappa = 1$. From the simulations, we can conclude that, for this scenario, during high demands, the MaxPressure controller performs better than the GPA controller, while during low demands the GPA performs better. One explanation for this could be that during low demands, adopting the cycle length is critical, while during high demands when almost all the sensors are covered, it is more important to keep the queue balanced between the current and downstream lanes. The proportional fair controller that does not adopt its cycle length, performs always the worst, and in most of the cases a fixed time controller performs second worst. It is just for the demand $\delta = 0.15$, and during the draining phase that the fixed time controller performs better than the GPA controller.

%\begin{figure}
%\centering
%\begin{tikzpicture}
%\begin{axis}[ymode=log, width=8cm,  height=6cm,  ylabel={Total Queue Length [m] }, xlabel={Time [s]}, xmax=6000,legend columns=2,  legend style={at={(0.5,-0.25)},anchor=north}]
%\addplot[mark=none, color=mycolor1] table [x index=0, y index=1]{plotdata/bpvspf/queue_pf2_k10_l0.05.csv};
%\addplot[mark=none, color=mycolor1, dashed] table [x index=0, y index=1]{plotdata/bpvspf/queue_bp_d10_l0.05.csv};
%\addplot[mark=none, color=mycolor2] table [x index=0, y index=1]{plotdata/bpvspf/queue_pf2_k10_l0.10.csv};
%\addplot[mark=none, color=mycolor2, dashed] table [x index=0, y index=1]{plotdata/bpvspf/queue_bp_d10_l0.10.csv};
%\addplot[mark=none, color=mycolor3] table [x index=0, y index=1]{plotdata/bpvspf/queue_pf2_k10_l0.15.csv};
%\addplot[mark=none, color=mycolor3, dashed] table [x index=0, y index=1]{plotdata/bpvspf/queue_bp_d10_l0.15.csv};
%\legend{GPA $\delta = 0.05$, MP $\delta = 0.05$,  GPA $\delta = 0.10$, MP $\delta = 0.10$,  GPA $\delta = 0.15$, MP $\delta = 0.15$  }
%\end{axis}
%\end{tikzpicture}
%\caption{Comparison between the GPA with $\kappa = 5$ and MaxPressure with $d =10$ for the three different demand levels in the Manhattan scenario. In order to improve the readability of the results, the queue-lengths are averaged over $300$ seconds interval.}
%\label{fig:comparison}
%\end{figure}

\begin{table}
\centering
\caption{Fixed Time and Proportional Fair Control - Manhattan Scenario}
\label{tab:fixedmanhattan}
\begin{tabular}{ccc}
Controller & $\delta$ & Total Travel Time [h] \\ \hline \hline
FT & $0.05$ & $1201$ \\
FT & $0.10$ & $2555$ \\
FT & $0.15$ & $4642$   \\
PF & $0.05$ & $1694$   \\
PF & $0.10$ & $4165$   \\
PF & $0.15$ & $+\infty$ \\ \hline
\end{tabular}
\end{table}

\begin{figure}
\centering
\begin{tikzpicture}
\begin{axis}[ymode=log, width=8cm,  height=6cm,  ylabel={Total Queue Length [m] }, xlabel={Time [s]}, xmax=5000, legend pos=north west, scaled y ticks = false,
      y tick label style={/pgf/number format/fixed,
      /pgf/number format/1000 sep = \thinspace % Optional if you want to replace comma as the 1000 separator 
      }, , legend style={at={(0.5,-0.25)},anchor=north}, legend columns=2]

\addplot[mark=none, color=mycolor1, thick] table [x index=0, y index=1]{plotdata/bpvspf/queue_pf2_k5_l0.05.csv};
\addplot[mark=none, color=mycolor2, thick] table [x index=0, y index=1]{plotdata/bpvspf/queue_pf2_k10_l0.05.csv};
\addplot[mark=none, color=mycolor3, thick] table [x index=0, y index=1]{plotdata/bpvspf/queue_bp_d10_l0.05.csv};
\addplot[mark=none, color=mycolor4, thick] table [x index=0, y index=1]{plotdata/bpvspf/queue_fixed_l0.05.csv};
\addplot[mark=none, color=black, thick] table [x index=0, y index=1]{plotdata/bpvspf/queue_pf_fixed_l0.05.csv};
\legend{GPA $\kappa =5$, GPA $\kappa =10$, MP $d = 10$, Fixed Time, PF}
\end{axis}
\end{tikzpicture}
\caption{A comparison between different control strategies for the Manhattan grid with the demand $\delta = 0.05$.o improve the readability of the results, the queue-lengths are averaged over $300$ seconds intervals.}
\label{fig:comparisonl0.05}
\end{figure}

\begin{figure}
\centering
\begin{tikzpicture}
\begin{axis}[ymode=log, width=8cm,  height=6cm,  ylabel={Total Queue Length [m] }, xlabel={Time [s]}, xmax=5500, legend pos=north west, scaled y ticks = false,
      y tick label style={/pgf/number format/fixed,
      /pgf/number format/1000 sep = \thinspace % Optional if you want to replace comma as the 1000 separator 
      }, , legend style={at={(0.5,-0.25)},anchor=north},  legend columns=2]

\addplot[mark=none, color=mycolor1, thick] table [x index=0, y index=1]{plotdata/bpvspf/queue_pf2_k5_l0.10.csv};
\addplot[mark=none, color=mycolor2, thick] table [x index=0, y index=1]{plotdata/bpvspf/queue_pf2_k10_l0.10.csv};
\addplot[mark=none, color=mycolor3, thick] table [x index=0, y index=1]{plotdata/bpvspf/queue_bp_d10_l0.10.csv};
\addplot[mark=none, color=mycolor4, thick] table [x index=0, y index=1]{plotdata/bpvspf/queue_fixed_l0.10.csv};
\addplot[mark=none, color=black, thick] table [x index=0, y index=1]{plotdata/bpvspf/queue_pf_fixed_l0.10.csv};
\legend{GPA $\kappa =5$, GPA $\kappa =10$, MP $d = 10$, Fixed Time, PF}
\end{axis}
\end{tikzpicture}
\caption{A comparison between different control strategies for the Manhattan grid with the demand $\delta = 0.10$. To improve the readability of the results, the queue-lengths are averaged over $300$ seconds intervals.}
\label{fig:comparisonl0.10}
\end{figure}

\section{LuST scenario} \label{sec:lust}

\begin{figure}
\centering
\includegraphics[width=0.48\textwidth]{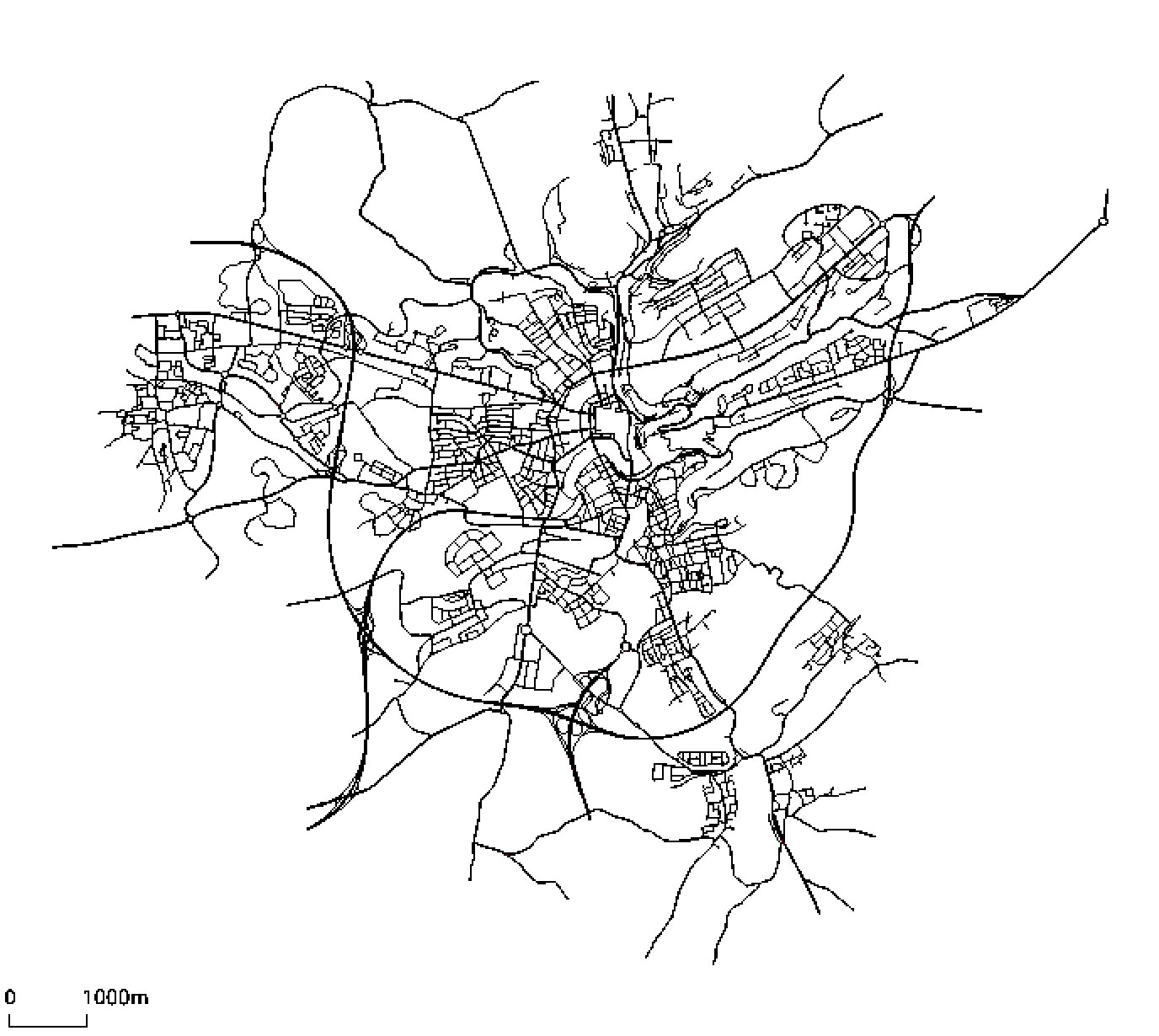}
\caption{The traffic network of Luxembourg city}
\label{fig:lust}
\end{figure}

To test the proposed controller in a realistic scenario, we make use of the Luxembourg SUMO Traffic (LuST) scenario presented in~\cite{codeca2017luxembourg}\footnote{The scenario files are obtained from \url{https://github.com/lcodeca/LuSTScenario/tree/v2.0}}. The scenario models the city center of Luxembourg during a full day, and the authors of~\cite{codeca2017luxembourg} have made several adjustments from some given population data when creating the scenario, to make it as realistic as possible.

The LuST network is shown in Fig.~\ref{fig:lust}. To each of the $199$ signalized junctions, we have added a lane area detector to each incoming lane. The length of the detectors are $100$ meters, or as long as the lane is if it is shorter than $100$ meters. Those sensors are added to give the controller real-time information about the queue-lengths at each junction.

As input to the system, we are using the Dynamic User Assignment demand data. For this data-set, the drivers try to take their shortest path (with respect to time) between their current position and destination. It is assumed that $70$ percent of the vehicles can recompute their shortest path while driving, and will do so every fifth minute. This rerouting possibility is introduced in order to model the fact that more and more drivers are using online navigation with real-time traffic state information, and will hence get updates about what the optimal route choice is.

In the LuST scenario, the phases are constructed in a bit more complex way and are not always orthogonal. For non-orthogonal phases, it is not always the case that all lanes receive yellow light when a clearance phase is activated. If the lane receives a green light in the next phase as well, it will receive green light during the clearance phase too. This property makes it more difficult to shorten the cycle, and for that reason, we choose to implement the controller which activates all the clearance phases in the cycle, i.e., the controller given in Section~\ref{sec:GPAfull}.

As mentioned, the phases in the LuST scenario are not orthogonal in each junction. Hence we have to solve the convex optimization problem in~\eqref{eq:gpa} to compute the phase activation. The computation is done by using the solver CVXPY\footnote{https://cvxpy.org} in Python. Although the controller can be implemented in a distributed manner, the simulations are in this paper performed on a single computer. Despite the size of the network, and that the communication via TraCI between the controller written in Python and SUMO slows down the simulations significantly, the simulations are still running about $2.5$ times faster than real-time.  This shows that there is no problem with running this controller in a real-time setting.

Since the demand is high during the peak-hours in the scenario, gridlock situations occur. Those kinds of situations is unavoidable since there will be conflicts in the car following model. To make the simulation continue to run, SUMO has a teleporting option that is utilized in the original LuST scenario. The original LuST scenario is configured such that if that a vehicle has been looked for more than $10$ minutes, it will teleport along its route until there is free space. It is therefore important when we evaluate the control strategies that we keep track of the number of teleports, to make sure that the control strategy will not create a significantly larger amount of gridlocks, compared to the original fixed time controller. In Table~\ref{tab:lust} the number of teleports are reported for each controller. It is also reported how many of those teleports that are caused directly due to traffic jam, but one should have in mind that e.g., a gridlock caused by that two vehicles want to swap lanes, is often a consequence of a congestion.

\begin{figure}
\begin{tikzpicture}
\begin{axis}[ymode=log, width=8cm,  height=8cm,  ylabel={Total Queue Length [m] }, xlabel={Time}, legend pos=north west, xmin=0, xmax=24.00, xtick={0, 4, 8, 12, 16, 20, 24},
x filter/.code={\pgfmathparse{#1/3600+0}},
xticklabel={ % Split into hours and minutes
        \pgfmathsetmacro\hours{floor(\tick)}%
        \pgfmathsetmacro\minutes{(\tick-\hours)*0.6}%
        % Use some trickery to get leading zeros
        \pgfmathprintnumber{\hours}:\pgfmathprintnumber[fixed, fixed zerofill, skip 0.=true, dec sep={}]{\minutes}%
    },
    legend columns=2,  legend style={at={(0.5,-0.25)},anchor=north}
]
\addplot[mark=none, color=mycolor1] table [x index=0, y index=1]{plotdata/csv/queue_pf_k10_tmin0.0.csv};
\addplot[mark=none, color=mycolor2] table [x index=0, y index=1]{plotdata/csv/queue_pf_k10_tmin0.1.csv};
\addplot[mark=none, color=mycolor3] table [x index=0, y index=1]{plotdata/csv/queue_pf_k10_tmin0.2.csv};
\addplot[mark=none, color=mycolor4] table [x index=0, y index=1]{plotdata/csv/queue_pf_k10_tmin0.30.csv};
\addplot[mark=none] table [x index=0, y index=1]{plotdata/csv/queue_pf_k10_tmin0.40.csv};
\addplot[mark=none, color=black, dotted] table [x index=0, y index=1]{plotdata/csv/queue_static.csv};
\legend{GPA  $\bar{w} = 0$, GPA  $\bar{w} = 0.1$, GPA $\bar{w} = 0.2$, GPA $\bar{w} = 0.3$, GPA $\bar{w} = 0.4$, Fixed Time }
\end{axis}
\end{tikzpicture}
\caption{How the queue lengths varies with time when the traffic lights in the LuST scenario are controlled with the GPA controller and the standard fixed time controller. For the GPA controller the paramters $\kappa = 10$ and different values of $\bar{w}$ are tested.  In order to improve the readability of the results, the queue-lengths are averaged over $300$ seconds intervals.}
\label{fig:lustkappa10}
\end{figure}
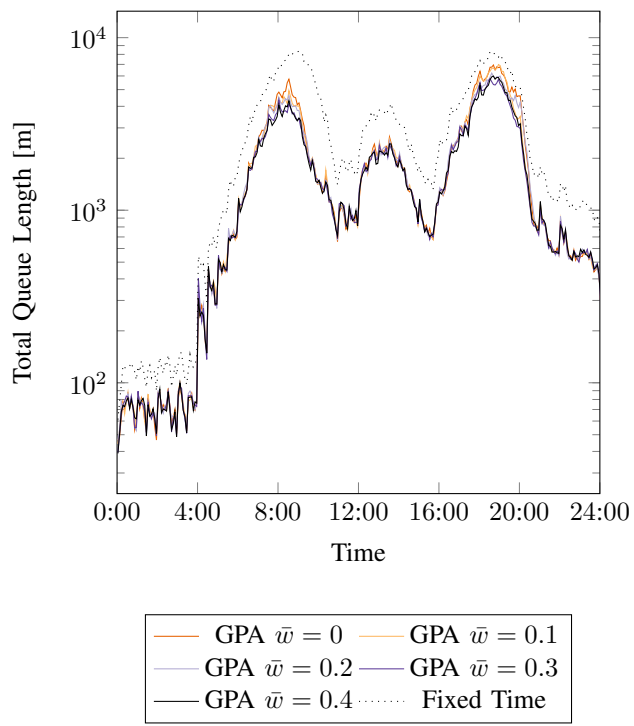

The total travel time and the number of teleports for different choices of tuning parameters are shown in Table~\ref{tab:lust}. For the fixed time controller, we keep the standard fixed time plan provided with the LuST scenario. How the queue lengths vary with time for different $\bar w$ is shown in Fig.~\ref{fig:lustkappa5} for $\kappa =5$ and in Fig.~\ref{fig:lustkappa10} for $\kappa = 10$.

From the results, we can see that any controller with $\kappa = 10$  and $\bar{w}$ within the range of investigation will improve the traffic situation. However, the controller that yields the overall shortest total travel time is the one with $\kappa =5$ and $\bar{w} = 0.40$. This result suggests that tuning the GPA only with respect to $\kappa$, and keep $\bar{w} = 0$, may not lead to the best performance with respect to total travel time, although it gives higher theoretical throughput.

\begin{table}
\caption{Comparison of the different control strategies}
\label{tab:lust}
\centering
\begin{tabular}{lcccc}
& $\kappa$ & $\bar{w}$ & Teleports (jam) & Total Travel Time [h] \\ \hline \hline
GPA & $10$ & $0$ & 76 (6) & 49 791 \\
GPA & $10$ & $0.05$ & 65 (1) & 49 708  \\
GPA & $10$ & $0.10$ & 37 (0) & 49 519 \\
GPA & $10$ & $0.15$ & 57 (19) & 49 408  \\
GPA & $10$ & $0.20$ & 50 (10) & 49 380 \\
GPA & $10$ & $0.25$ & 35 (0) &  49 265\\
GPA & $10$ & $0.30$ & 30 (0) &  48 930\\
GPA & $10$ & $0.35$ & 25 (1) &  48 922\\
GPA & $10$ & $0.40$ & 51 (0) &  48 932 \\
GPA & $10$ & $0.45$ & 49 (5) &  49 076 \\
GPA & $10$ & $0.50$ & 42 (15) & 49 383 \\
GPA & $5$ & $0$ & 668 (76) & 57 249 \\
GPA & $5$ & $0.05$ & 234 (62) & 54 870  \\
GPA & $5$ & $0.10$ & 68 (10)  & 52 038 \\
GPA & $5$ & $0.15$ & 47 (9)  & 50 696 \\
GPA & $5$ & $0.20$ & 50 (6) & 49 904 \\
GPA & $5$ & $0.25$ & 41 (3) & 49 454 \\
GPA & $5$ & $0.30$ & 23 (0) & 48 964 \\
GPA & $5$ & $0.35$ & 30 (1) & 48 643 \\
GPA & $5$ & $0.40$ & 35 (5) & 48 445 \\
GPA & $5$ & $0.45$ & 39 (1) & 48 503 \\
GPA & $5$ & $0.50$ & 42 (10) & 48 772 \\
Fixed time & -- & -- & 122 (80) & 54 103\\ \hline
\end{tabular}

\end{table}

\begin{figure}
\begin{tikzpicture}
\begin{axis}[ymode=log, width=8cm,  height=8cm,  ylabel={Total Queue Length [m] }, xlabel={Time}, legend pos=north west, xmin=0, xmax=24.00, xtick={0, 4, 8, 12, 16, 20, 24},
x filter/.code={\pgfmathparse{#1/3600+0}},
xticklabel={ % Split into hours and minutes
        \pgfmathsetmacro\hours{floor(\tick)}%
        \pgfmathsetmacro\minutes{(\tick-\hours)*0.6}%
        % Use some trickery to get leading zeros
        \pgfmathprintnumber{\hours}:\pgfmathprintnumber[fixed, fixed zerofill, skip 0.=true, dec sep={}]{\minutes}%
    },
    legend columns=2,  legend style={at={(0.5,-0.25)},anchor=north}
]
\addplot[mark=none, color=mycolor1] table [x index=0, y index=1]{plotdata/csv/queue_pf_k5_tmin0.0.csv};
\addplot[mark=none, color=mycolor2] table [x index=0, y index=1]{plotdata/csv/queue_pf_k5_tmin0.1.csv};
\addplot[mark=none, color=mycolor3] table [x index=0, y index=1]{plotdata/csv/queue_pf_k5_tmin0.2.csv};
\addplot[mark=none, color=mycolor4] table [x index=0, y index=1]{plotdata/csv/queue_pf_k5_tmin0.30.csv};
\addplot[mark=none] table [x index=0, y index=1]{plotdata/csv/queue_pf_k5_tmin0.40.csv};
\addplot[mark=none, color=black, dotted] table [x index=0, y index=1]{plotdata/csv/queue_static.csv};
\legend{GPA  $\bar{w} = 0$, GPA  $\bar{w} = 0.1$, GPA $\bar{w} = 0.2$, GPA $\bar{w} = 0.3$, GPA $\bar{w} = 0.4$, Fixed Time }
\end{axis}
\end{tikzpicture}
\caption{How the queue lengths varies with time when the traffic lights in the LuST scenario are controlled with the GPA controller and the standard fixed time controller. For the GPA controller the paramters $\kappa = 5$ and different values of $\bar{w}$ are tested.  In order to improve the readability of the results, the queue-lengths are averaged over $300$ seconds intervals.}
\label{fig:lustkappa5}
\end{figure}

\section{Conclusions}

In this paper, we have discussed implementational aspects of the Generalized Proportional Allocation controller. The controller's performance was compared to the MaxPressure controller both on an artificial Manhattan-like grid and for a real scenario. In comparison with MaxPressure, it was shown that the controller performs better than the MaxPressure controller when the demand is low, but the MaxPressure performs better during high demand. Those observations hold true even if the MaxPressure controller does not have correct information about the turning ratios in each junction.

While the information about the turning ratios and the queue lengths at neighboring junctions are needed for the MaxPressure controller, the GPA controller does not require any such information. This makes the GPA controller easier to implement in a real scenario, where the downstream junction may not be signalized and equipped with sensors. We showed that it is possible to both implement the GPA controller in a realistic scenario covering the city of Luxembourg and that it improves the traffic situation compared to a standard fixed time controller.

In all simulations, we have used the same tuning parameters for all junctions in the LuST scenario, while the fixed time controller is different for different junction settings. Hence the GPA controller's performance can be even more improved by tuning the parameters specifically for each junction. Ideally, this should be done with some auto-tuning solution, but it may also be worth to take static parameters into account, such as the sensor lengths. This is a topic for future research.

\bibliographystyle{ieeetr}%
\bibliography{ref}             

%\begin{IEEEbiography}[{\includegraphics[width=1in,he ight=1.25in,clip,keepaspectratio]{./shell}}]{Michael Shell}
\newpage
\begin{IEEEbiography}[{\includegraphics[width=1in,height=1.25in,clip,keepaspectratio]{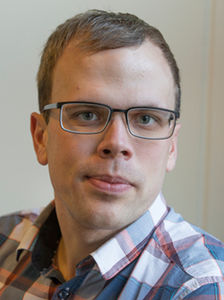}}]{Gustav Nilsson} received his Master of Engineering Physics degree from Lund University in 2013. Since then, he has been a PhD student at the Department of Automatic Control, Lund University, working with Prof. Giacomo Como. During his PhD studies, he has done longer research visits to Institute of Pure and Applied Mathematics (IPAM), UCLA, CA, USA and  Department of Mathematical Sciences, Politecnico di Torino, Turin, Italy. Between October 2017 and March 2018, he did an internship at Mitsubishi Electric Research Laboratories in Cambridge, MA, USA. His primary research interest lies in modeling and control of dynamical flow networks with applications in transportation networks.\end{IEEEbiography}
\vspace{-120 mm}
\begin{IEEEbiography}[{\includegraphics[width=1in,height=1.25in,clip,keepaspectratio]{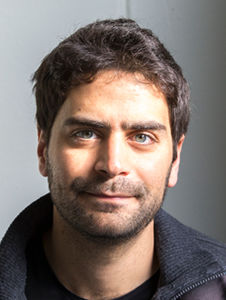}}]{Giacomo Como}  is an Associate Professor at the Department of Mathematical Sciences, Politecnico di Torino, Italy, and at the Automatic Control Department of Lund University, Sweden. He received the B.Sc., M.S., and Ph.D. degrees in Applied Mathematics from Politecnico di Torino, in 2002, 2004, and 2008, respectively. He was a Visiting Assistant in Research at Yale University in 2006-2007 and a Postdoctoral Associate at the Laboratory for Information and Decision Systems, Massachusetts Institute of Technology, from 2008 to 2011. He currently serves as Associate Editor of IEEE-TCNS and IEEE-TNSE and as chair of the IEEE-CSS Technical Committee on Networks and Communications. He was the IPC chair of the IFAC Workshop NecSys'15 and a semiplenary speaker at the International Symposium MTNS'16. He is recipient of the 2015 George S. Axelby Outstanding Paper award. His research interests are in dynamics, information, and control in network systems with applications to cyber-physical systems, infrastructure networks, and social and economic networks.
\end{IEEEbiography}

\end{document}